\documentclass{aa}  

\usepackage{txfonts}
\usepackage{graphicx}
\usepackage{lscape}
\usepackage{amsfonts,amsmath,natbib,xcolor,subfigure,gensymb,longtable,ulem}

\defcitealias{Golkhou14}{GB14}
\defcitealias{Golkhou15}{GBL15}

\begin{document} 

\title{GRB minimum variability timescale with {\it Insight-HXMT} and {\it Swift}:}
\subtitle{implications for progenitor models, dissipation physics and GRB classifications}

\author{A.~E.~Camisasca\thanks{annaelisa.camisasca@unife.it}\inst{1}
        \and C.~Guidorzi\inst{1,2,3} 
        \and L.~Amati \inst{3}
        \and F.~Frontera\inst{1,3}
        \and X.Y.~Song\inst{4}
        \and S.~Xiao\inst{4,5}
        \and S.L.~Xiong\inst{4}
        \and S.N.~Zhang\inst{4,5}
        \and R.~Margutti \inst{6,7}
        \and S.~Kobayashi \inst{8}
        \and C.G.~Mundell \inst{9}
        \and M.Y.~Ge \inst{4}
        \and A.~Gomboc \inst{10}
        \and S.M.~Jia \inst{4}
        \and N.~Jordana-Mitjans \inst{9}
        \and C.K.~Li \inst{4}
        \and X.B.~Li \inst{4}
        \and R.~Maccary \inst{1}
        \and M.~Shrestha \inst{11}
        \and W.C.~Xue \inst{4}
        \and S.~Zhang \inst{4}
                       }

\institute{Department of Physics and Earth Science, University of Ferrara, via Saragat 1, I--44122, Ferrara, Italy
    \and INFN -- Sezione di Ferrara, via Saragat 1, I--44122, Ferrara, Italy
    \and INAF -- Osservatorio di Astrofisica e Scienza dello Spazio di Bologna, Via Piero Gobetti 101, I-40129 Bologna, Italy
    \and Key Laboratory of Particle Astrophysics, Institute of High Energy Physics, Chinese Academy of Sciences, 19B Yuquan Road, Beijing 100049, People’s Republic of China
    \and University of Chinese Academy of Sciences, Chinese Academy of Sciences, Beijing 100049, China
    \and Department of Astronomy, University of California, 501 Campbell Hall, Berkeley, CA 94720, USA
    \and Department of Physics, University of California, 366 Physics North MC 7300, Berkeley, CA 94720, USA
    \and  Astrophysics Research Institute, Liverpool John Moores University, IC2, Liverpool Science Park, 146 Brownlow Hill, Liverpool L3 5RF, UK
    \and Department of Physics, University of Bath, Claverton Down, Bath, BA2 7AY, UK
    \and Center for Astrophysics and Cosmology, University of Nova Gorica, Vipavska 13, 5000 Nova Gorica, Slovenia
    \and Steward Observatory, University of Arizona, 933 North Cherry Avenue, Tucson, AZ 85721-0065, USA
    }

\date{}

\abstract
{There has been significant technological and scientific progress in our ability to detect, monitor and model the physics of gamma--ray bursts (GRBs) over the 50 years since their first discovery. However, the dissipation process thought to be responsible for their defining prompt emission is still unknown. Recent efforts have focused on investigating how the ultrarelativistic jet of the GRB propagates through the progenitor’s stellar envelope, for different initial composition shapes, jet structures, magnetisation, and –- consequently –- possible energy dissipation processes. Study of the temporal variability -- in particular the shortest duration of an independent emission episode within a GRB -- may provide a unique way to discriminate the imprint of the inner engine activity from geometry and propagation related effects. The advent of new high-energy detectors with exquisite time resolution now makes this possible.}
{We aim to characterise the minimum variability timescale (MVT) defined as the shortest duration of individual pulses that shape a light curve for a sample of GRBs in the keV–MeV energy range and test correlations with other key observables, such as the peak luminosity, the Lorentz factor, and the jet opening angle. We compare these correlations with predictions from recent numerical simulations for a relativistic structured -- possibly wobbling -- jet and assess the value of temporal variability studies as probes of prompt-emission dissipation physics.}
{We used the peak detection algorithm {\sc mepsa} to identify the shortest pulse within a GRB time history and preliminarily calibrated {\sc mepsa} to estimate the full width half maximum (FWHM) duration. We then applied this framework to two sets of GRBs: {\it Swift} GRBs (from 2005 to July 2022) and Insight Hard Modulation X-ray Telescope ({\it Insight-HXMT}) GRBs (from June 2017 to July 2021, including the exceptional 221009A). We then selected 401 GRBs with measured redshift to test for correlations.} 
{We confirm that on average short GRBs have significantly shorter MVT than long GRBs. The MVT distribution of short GRBs with extended emission such as 060614 and 211211A is compatible only with that of short GRBs. This is important because it provides a new clue on the progenitor's nature. The MVT for long GRBs with measured redshift anti-correlates with peak luminosity; our analysis includes careful evaluation of selection effects. We confirm the anti-correlation with the Lorentz factor and find a correlation with the jet opening angle as estimated from the afterglow light curve, along with an inverse correlation with the number of pulses.}
{The MVT can identify the emerging putative new class of long GRBs that are suggested to be produced by compact binary mergers. For otherwise typical long GRBs, the different correlations between MVT and peak luminosity, Lorentz factor, jet opening angle, and number of pulses can be explained within the context of structured, possibly wobbling, weakly magnetised relativistic jets.}


\maketitle

%

\section{Introduction}
\label{par:intro}
The prompt emission of gamma--ray bursts (GRBs) is the first energetic and short-lived electromagnetic messenger produced by a relativistic jet that forms in at least two classes of progenitors:  (i) binary compact object mergers, where at least one of the two components is thought to be a neutron star (NS; \citealt{Eichler89,Paczynski91,Narayan92}); (ii) collapsars, massive stars whose core collapses to a compact object, which powers a relativistic jet that breaks out of the stellar envelope \citep{Woosley93,Paczynski98,MacFadyen99}. Most GRBs of the former (latter) class have a typical duration of a few $0.1$~s ($>$ several seconds), and are therefore referred to as "short" ("long"), hereafter SGRBs (LGRBs). The discovery of so-called soft extended emission short GRBs (hereafter SEE-SGRBs; \citealt{Norris06}), i.e. GRBs whose duration is formally long ($T_{90}>2$~s, usually taken as the boundary), but whose profile includes an initial hard subsecond spike followed by a several-second-long-lived soft tail, and for which evidence points to class (i), emphasised that duration alone can be occasionally misleading (e.g. \citealt{Amati21}). The occurrence of baffling SEE-SGRBs, such as 060614 \citep{Gehrels06,DellaValle06,Fynbo06,Jin15} and the recent 211211A \citep{Rastinejad22,Gompertz22,Yang22,Troja22,Xiao22} shows that events of class (i) may exhibit time profiles that further challenge and elude this picture. Opposite cases of apparently SGRBs that instead belong to (ii) have also been identified \citep{Ahumada21,Zhang21a,Rossi22a}, so that in the hindsight the emerging picture raises the issue of a contamination between the two classes that is potentially more widespread than what was thought so far, purely based on time profiles. Consequently, to avoid confusion and adopting previous suggestions \citep{Zhang06_nat,Zhang07b,Kann11,KWGRBcat17}, when talking about progenitor case, hereafter we will refer to (i) and (ii) candidates as Type-I and Type-II, respectively.

Many open and intertwined issues still enshroud the GRB prompt emission: which source of energy turns into gamma-rays, bulk kinetic or magnetic energy? Ruled by which dissipation process? What is the composition of the relativistic jet? At which distance from the inner progenitor does the dissipation take place? Among the distinctive properties are the variety and complexity of GRB light curves (LCs), which manifest as a wide range of variability over several timescales. While this complexity likely retains a wealth of information, a full understanding has yet to be found.

This variability can be the result of several different contributions: (a) inner engine activity both in terms of short timescales and number of peaks \citep{Kobayashi97}; (b) propagation of the relativistic flow through the stellar envelope, which in turn also depends on (c) the jet composition \citep{Gottlieb19,Gottlieb20a,Gottlieb20b,Gottlieb21a,Gottlieb21b}; (d)  geometry: structure of the jet and angle of the observer, $\theta_{\rm obs}$, with respect to the opening angle of the jet core, $\theta_j$ (e.g., \citealt{Salafia16}); (e) a possible precessing/wobbling jet \citep{PortegiesZwart99,PortegiesZwart01,Fargion01,Reynoso06,Lei07,Budai20}, as suggested by state-of-the-art 3-D GRMHD simulations \citep{Gottlieb22b}.

Numerous definitions of GRB variability have been put forward in the literature, aimed to quantify the net variance of the GRB signal, once the contribution of the noise due to counting statistics (hereafter statistical noise) is removed. This was done either summing the contributions of all timescales after excluding some kind of trend (e.g., \citealt{Reichart01,Fenimore00}), or decomposing the signal variance over a timescale base. In the latter case, the decomposition can be done either in time \citep{Li01,Margutti09,Margutti11c} or frequency domain, based on either Fourier analysis \citep{Guidorzi16,Dichiara16} or wavelets (\citealt[hereafter GB14]{Golkhou14}; \citealt{Golkhou15,Vianello18}).
In spite of the scatter, for LGRBs variability was found to correlate with peak luminosity \citep{Fenimore00,Reichart01,Guidorzi05b}, although the slope can vary remarkably depending on the definition of variability and other aspects \citep{Guidorzi06b}.

A related way to characterise the variability is the identification of the minimum variability timescale (MVT)\footnote{Shortened to "MTS" in some papers.}, which is the shortest timescale at which uncorrelated power is found significantly in excess of the statistical noise as well as of the correlated signal due to the overall temporal shape of the GRB LC. In principle this quantity helps to constrain the emitting region size and the activity of the inner engine and could help identify the different contributions (a)--(e) listed above, especially when it is studied in conjunction with other key properties.

The two classes of SGRBs and LGRBs have partially overlapping, but different MVT distributions, with median rest-frame values of 10 and 45~ms, respectively, and very few ($<10$\%) with ms MVT (\citealt{Golkhou15}; see also \citealt{MacLachlan13}).

The MVT of LGRBs was found to correlate with the bulk Lorentz factor $\Gamma_0$ \citep{Sonbas15} estimated from the early afterglow peak, whenever this is due to the deceleration of the relativistic jet in the thin shell regime \citep{Sari99,Molinari07}.

A drawback of most definitions relies in the meaning itself and how this is to be interpreted: while connections are sometimes found with simple properties, such as the individual pulse rise time \citep{MacLachlan12}, the interpretation is not straightforward. This is partly due to the complexity itself of the GRB signal, which is short-lived, highly non-stationary and occasionally with an evolving power density spectrum (e.g., \citealt{Margutti08b}). A common feature of most definitions of MVT is that the identification of one or more temporal structures associated with MVT relies on their relative weight in the total net variance of the GRB LC. As a consequence, a given spike could be identified or not, depending on its impact on the time-averaged power density spectrum.

In this paper, we adopted an alternative approach that builds on a simple definition of MVT as the full-width half-maximum (FWHM) of the shortest (statistically significant) peak (hereafter, FWHM$_{\rm min}$). The identification of statistically significant peaks is done using the sensitive and well-calibrated algorithm {\sc mepsa} \citep{Guidorzi15a}. A similar idea, based on the identification of individual pulses within a GRB, was already proposed in the past \citep{Bhat12,Bhat13}, but was not explored any further. This approach has three main advantages: 1) the interpretation is straightforward; 2) it is related directly to a specific temporal structure within the overall GRB time profile and, as such, the probability of being identified does not depend on its relative weight within the total variance of the GRB; 3) a careful evaluation of how the measure of MVT is affected by the signal-to-noise ratio (SNR) is feasible and, consequently, of the impact on the correlations involving MVT.

To this aim, we carried out our analysis using two GRB catalogues with complementary energy passbands: the first GRB catalogue \citep{Song22_HXMTGRBcatalog} of the Insight Hard X-ray Modulation Telescope ({\it Insight-HXMT}; \citealt{Li07,Zhang20_HXMT}) and that of the Burst Alert Telescope (BAT, 15-150 keV; \citealt{Barthelmy05}) aboard the {\it Neil Gehrels Swift Observatory} \citep{Gehrels04}.

The {\it HXMT}, named ``Insight'' after launch on June 15, 2017, is the first Chinese X--ray astronomy satellite. Its scientific payload consists of three main instruments: the  Low Energy X--ray telescope (LE; 1--15~keV; \citealt{Chen20_HXMT}), the Medium Energy X--ray telescope (ME; 5--30~keV; \citealt{Cao20_HXMT}), and the High Energy X--ray telescope (HE; \citealt{Liu20_HXMT}). The HE consists of 18 NaI/CsI detectors which cover the 20--250~keV energy band for pointing observations. In addition, it can be used as an open sky GRB monitor in the $0.2$--3~MeV energy range. The unique combination of a large geometric area ($\sim5100$\,cm$^{2}$) and of continuous event tagging with timing accuracy $<10\,\mu$s, makes {\it HXMT}/HE an ideal instrument to study MVTs with GRBs.
In this work we investigate this possibility by carrying out a systematic analysis of the data acquired with HE, used as an open sky $\gamma$--ray monitor.

Section~\ref{sec:dataset} describes the GRB samples; the data analysis is reported in Section~\ref{sec:data_an}, whereas results are in Section~\ref{sec:res}. We discuss the implications in Section~\ref{sec:disc} and conclude in Section~\ref{sec:conc}. Hereafter, we assume the latest Planck cosmological parameter
s: $H_0=67.74$~km\,s$^{-1}$\,Mpc$^{-1}$, $\Omega_m=0.315$, $\Omega_\Lambda=0.685$ \citep{cosmoPlanck20}.

\section{Data set}
\label{sec:dataset}

\subsection{{\it Swift}/BAT sample}
\label{subsec:BATset}
We considered all GRBs detected by {\it Swift}/BAT in burst mode from January 2005 to July 2022. We extracted the mask-weighted LCs in the 15--150~keV energy band following the standard procedure recommended by the BAT team\footnote{\url{https://swift.gsfc.nasa.gov/analysis/threads/bat_threads.html}} with a set of different uniform bin times: 1, 4, 64, and 1000 ms. We excluded all the GRBs whose LCs were not entirely covered in burst mode. We then systematically applied {\sc mepsa} to each LCs. For each GRB the FWHM$_{\rm min}$ was determined through the procedure described in Section~\ref{sec:data_an}. Finally, we ignored the GRBs for which either FWHM$_{\rm min}$ or $T_{90}$ is not significant. The final sample of GRBs includes 1291~GRBs ({\it Swift} sample hereafter). For 21 GRBs only an upper limit on the FWHM$_{\rm min}$ was derived. We also considered the duration of each GRB (expressed in terms of $T_{90}$) using the values by \citet{Lien16}, which cover up to October 2015. For the remaining GRBs we adopted the values reported by the BAT team through dedicated BAT refined Gamma-ray Coordinates Network (GCN) circulars.\footnote{\url{https://gcn.gsfc.nasa.gov/gcn3_archive.html}.}
For about one third of the sample (401/1291) the redshift is known. 

Although a lower threshold was suggested for {\it Swift} \citep{Bromberg13}, we take the value of 2~s as approximate boundary between S- and LGRBs, in line with the traditional division \citep{Kouveliotou93} and with the choice of {\it Swift} team members \citep{Davanzo14}. In this way we have 78 SGRBs. In addition, the sample includes 24 SEE-GRBs.\footnote{They are: 050724, 051227, 060614, 061006, 061210, 070714B, 071227, 080503, 090510, 090531B, 090715A, 090916, 110402A, 111121A, 150424A, 160410A, 161129A, 170728B, 180618A, 180805B, 181123B, 200219A, 211211A, 211227A.}
The overall sample consists of 102 Type-I GRBs, i.e. 8\% of the whole sample. Table~\ref{tab:BATsample} reports the data.

\begin{table*}[]
    \centering
    \begin{tabular}{l|c|c|c|c|c}
    \hline
    \hline
     GRB Name&FWHM$_{\rm min}$ (s) & $T_{90}$ (s) & $z$ &$N_{\rm peaks}$ & Type\\
     \hline
     \hline
      & & & & & \\
 050117&$0.810_{\tiny{-0.208}}^{\tiny{+0.280}}$&  $166.648 \pm 2.423$& -& $15$&L\\[0.2cm]
050124& $1.009_{-0.259}^{+0.349}$ &   $3.936 \pm 2.012$ & -&$2$&L\\[0.2cm]
 050126&$9.063_{-2.329}^{+3.134}$&$48.000 \pm 22.627$& $1.290$&$1$&L\\[0.2cm]
050128&$0.296_{-0.076}^{+0.102}$&$28.000 \pm 9.055$& -&$7$&L\\[0.2cm]
 050202&$ \leq 0.103_{-0.026}^{+0.036}$& $0.112 \pm 0.031$& -&$1$&S\\[0.2cm]
 050215A&$6.238_{-1.603}^{+2.158}$&$66.412 \pm 5.307$& -&$1$&L\\ [0.2cm]
 050215B&$4.384_{-1.127}^{+1.516}$&$11.044 \pm 3.931$& -&$1$&L\\ [0.2cm]
 050219A&$7.432_{-1.910}^{+2.570}$&$23.812 \pm 2.258$& $0.211$&$1$&L\\ [0.2cm]
050219B&$0.770_{-0.198}^{+0.266}$&$28.720 \pm 7.120$& -&$4$&L\\ [0.2cm]
050223&$23.518_{-6.044}^{+8.134}$&$22.680 \pm 4.481$& $0.592$&$1$&L\\ [0.2cm]

    \hline
    \end{tabular}
    \caption{The first 10 GRBs of {\it Swift} sample; this table is available in its entirety in machine-readable form.}
    \label{tab:BATsample}
\end{table*}

\begin{table*}[]
    \centering
    \begin{tabular}{l|c|c|c|c|c}
    \hline
    \hline
     GRB Name& FWHM$_{\rm min}$ (s) &$T_{90}$ (s)&$z$& $N_{\rm peaks}$&Type\\
     \hline
     \hline
      & & & & & \\

170626A & $  0.097_{   -0.025}^{    +0.033}$ &   $12.690 \pm   0.081$ &- &  4 &  L \\ [0.2cm]
170626B & $  1.969_{   -0.506}^{    +0.681}$ &   $6.511  \pm  1.120$ &-  & $2$ &  L \\ [0.2cm]
170705A & $  0.784_{   -0.201}^{    +0.271}$ &  $18.460  \pm  6.340$ &   $2.010$  & $1$ &  L \\ [0.2cm]
170708A & $  0.041_{   -0.011}^{    +0.014}$ &   $0.200  \pm 0.022$ &-  & $1$ &  S \\ [0.2cm]
170712A & $  1.188_{   -0.305}^{    +0.411}$ &   $8.511  \pm  1.119$ &-  & $1$ &  L \\ [0.2cm]
170718A & $  6.802_{   -1.748}^{    +2.353}$ &  $24.160 \pm 3.322$ &-  & 1 &  L \\ [0.2cm]
170726A & $  0.149_{   -0.038}^{    +0.051}$ &  $22.871 \pm 0.901$ &-  & $8$ &  L \\ [0.2cm]
170728B & $ \leq 0.275_{   -0.071}^{    +0.095}$ &  $16.860 \pm 2.371$ &-  & $2$ &  L \\ [0.2cm]
170801A & $  0.021_{   -0.005}^{    +0.007}$ &   $0.460 \pm 0.750$ &-  & $1$ &  S \\ [0.2cm]
170802A & $  0.033_{   -0.009}^{    +0.012}$ &   $0.820 \pm 0.014$ &-  & $2$ &  S \\ [0.2cm]
    \hline
    \end{tabular}
    \caption{The first 10 GRBs of {\it Insight-HXMT} sample; this table is available in its entirety in machine-readable form.}
    \label{tab:HXMTsample}
\end{table*}

\subsection{{\it Insight-HXMT}/HE sample}
\label{subsec:HXMVTet}
We considered all GRBs detected by {\it Insight-HXMT}/HE from June 2017 to June 2021, as catalogued by the {\it Insight-HXMT} team \citep{Song22_HXMTGRBcatalog}.
Since HE continuously acquires data in event mode, it has no trigger logic on board. For each GRB, whenever the GRB was detected in common by other experiments, such as {\it Swift}/BAT or {\it Fermi}/GBM, we took as the GRB start the trigger time provided by them. Differently, we determined the start time by visual inspection of the HE LC.

For each GRB we extracted the event files and auxiliary files including time-resolved information about the detectors’ dead time, spacecraft’s attitude and position, within a time window from $-300$ to $300$~s around the GRB time. Using the HE units as an open-sky monitor, for each of the 18 HE detectors, we extracted a set of LCs with the same bin times as for the {\it Swift} sample (i.e., 1, 4, 64, and 1000~ms) selecting only the CsI events, as was done in \citet{Song22_HXMTGRBcatalog}. The LCs include dead-time corrected counts within the total energy passband from all 18 HE detectors summed together. The total energy passband depends on the HE operation mode:
\begin{itemize}
\item normal mode: 80--800 keV;
\item GRB mode (low gain): 200--3000 keV.
\end{itemize}

We analysed separately 21 GRBs\footnote{They are: 171011B, 180113B, 180113C, 180218A, 180720B, 180914B, 181222B, 190103A, 190114C, 190305A, 190411A, 190415A, 190530A, 190606A, 190706C, 190928A, 191025B, 191227B, 201016A, 201227A, and 221009A. 200415A saturated the electronics too, but was not considered, as it is probably an extra-galactic magnetar giant flare; \citealt{Yang20,LAT21,Roberts21,Svinkin21}. }: due to their very intense peak count rates, the onboard electronics of at least one Physical Data Acquisition Units (PDAU, see \citealt{Liu20_HXMT} for details) were temporarily unable to keep up with the exceptional rate of events to be recorded (see \citealt{Xiao20,Song22_HXMTGRBcatalog} for details). We restricted our analysis to the time windows where no PDAU was saturated. For these GRBs we consequently ended up with upper limits on the FWHM$_{\rm min}$. In the following, we will consider the LCs summed over the 18 detectors.

The background was estimated in two independent ways: (i) through interpolation with a up to a third-degree polynomial within two time windows, preceding and following the GRB respectively. The size of each time window varies for different GRBs and had to be determined by visual inspection; (ii) by iterative interpolation of a unique time interval that includes both time windows used in (i) as well as the GRB interval: at every iteration all the time bins whose counts exceeded by $\ge 2\sigma$ the interpolated signal, where $\sigma$ is the corresponding Poisson uncertainty, are rejected. Iterations stop when no further bins are rejected. This iterative procedure was applied to the 1-s LC and the resulting background model was then properly renormalised to the LCs with different bin times. To determine which of the two outcomes is to be used for each GRB, we calculated the null hypothesis probability (NHP) associated with a two-tail $\chi^2$ test applied to the residuals of each LC with respect to each background model and chose the more probable one, provided that NHP was $\ge1$\%.

With the exception of the saturated GRBs, we used the $T_{90}$ values reported for the GRBs belonging to golden and silver samples of the {\it HXMT} GRB catalogue \citep{Song22_HXMTGRBcatalog}. For the saturated sample, we used the $T_{90}$ as reported on {\it Konus-Wind}, {\it Fermi-GBM}, {\it Insight-HXMT} GCN circulars. If different estimates of $T_{90}$ were provided by different experiments for a given GRB, we conservatively used mean and error that include all the values.

We decided to include the recent exceptionally bright 221009A \citep{Dichiara22_221009A}. Since it repeatedly saturated the electronics of {\it Insight-HXMT}/HE \citep{Ge22_221009A}, we provided an upper limit on the FWHM$_{\rm min}$ as for the other saturated GRBs. We estimated $T_{90}$ using the data of {\it BepiColombo}-MGNS in the 280--460 keV passband, which has one of the few unsaturated and publicly available time profiles \citep{BepiColombo22_221009A}. The 2-s time resolution is too coarse to constrain the MVT, but is enough for the $T_{90}$, given the very long duration. The background was interpolated linearly using the intervals -900 s $\leq$ t $\leq$ 100 s and 670 s $\leq$ t $\leq$ 1600 s, with $t$ measured since 13:15:26.90~UTC.

Finally, we systematically ran {\sc mepsa} on all the LCs and applied the procedure described in Section~\ref{sec:data_an} to determine the FWHM$_{\rm min}$ of each GRB as we did for the {\it Swift} sample (Section~\ref{subsec:BATset}). Unlike for the BAT LCs, for which the Gaussian-noise regime is ensured by the fact that the rates are linear combinations of several thousands of independent detectors, in the case of the 1-ms LCs summed over all the 18 HE detectors the mean counts per bin amount to $\lesssim10$. Hence, the Gaussian-noise assumption is only approximately matched. To partially account for this deviation and use a more conservative estimate of the variance of a Poisson variate for small numbers, we corrected the uncertainties following the prescriptions by \citet{Gehrels86}.

We ignored the GRBs having non significant values for either FWHM$_{\rm min}$ or $T_{90}$, as we did with the {\it Swift} sample. The final sample of {\it Insight-HXMT}/HE includes 212~GRBs, 25 of which were detected in GRB mode, while the remaining ones in normal mode. Hereafter we will refer to them as the {\it HXMT} sample. For 14 GRBs only an upper limit on the FWHM$_{\rm min}$ was obtained. Taking the value of 2~s as an approximate boundary between SGRBs and LGRBs, we find that 24 are SGRBs, while 2 are SEE-GRBs. Thus 26 are Type-I GRBs, which correspond to 12\% of the total. This result shows that {\it Insight-HXMT} is more effective in detecting short hard GRBs than {\it Swift}  (7\%; Section~\ref{subsec:BATset}).

For 6 GRBs the redshift is known. There are 44 GRBs that were detected by {\it Swift}/BAT and {\it Insight-HXMT} and for which a comparative analysis of our results is feasible and done in section~\ref{sec:FWHM_vs_energy}. Data are reported in Table~\ref{tab:HXMTsample}.

\section{Data analysis}
\label{sec:data_an}
We applied {\sc mepsa} to the dataset described in Section~\ref{sec:dataset} in order to obtain to FWHM$_{\rm min}$.
In particular, for each GRB, we started from the 64-ms LC. A detected peak is considered a candidate when it satisfies the following two conditions: (i) SNR$> {\rm SNR}_{\rm min}^{(\Delta t)}$, where the threshold depends on the LC bin time $\Delta t$; (ii) $\Delta t_{\rm det} > \Delta t_{\rm det,min}^{(\Delta t)}$, where $\Delta t_{\rm det}$ is the so-called detection timescale, a {\sc mepsa} parameter which defines the bin timescale that optimises the peak identification \citep{Guidorzi15a}. Both sets of thresholds are reported in Table \ref{tab:threshold}.
The requirement (ii) is to ensure that the bin time is short enough to resolve the temporal structure: the duration of the peak must be greater than the size of the bin. This is the reason for setting the threshold to twice the corresponding bin time: $\Delta t_{\rm det,min}^{(\Delta t)} = 2 \Delta t$. The different threshold values on the SNR for different $\Delta t$ were calculated to keep approximately constant the number of expected statistical fluxes being classified as genuine peaks: the shorter $\Delta t$ and the correspondingly larger the number of bins to be screened that span a given time interval.

\begin{figure*}
\centering
\includegraphics[width=15 cm]{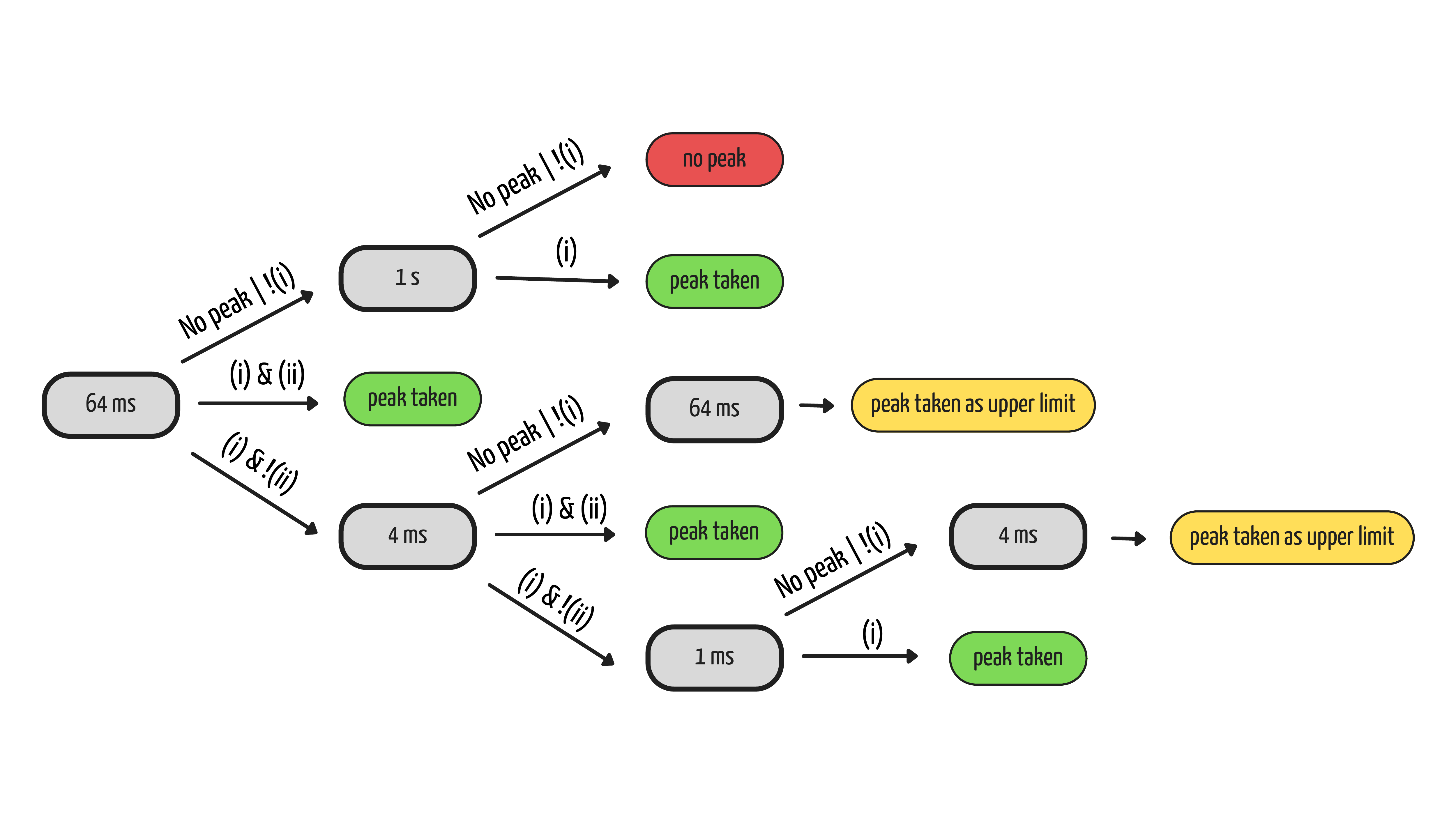}
\caption{A schematic description of the procedure adopted to determine the FWHM$_{\rm min}$ of each GRB (see Section~\ref{sec:data_an}).}
\label{fig:scheme}
\end{figure*}

We decided to start with $\Delta t=64$~ms and avoided a systematically research with finer resolution because preliminary attempts showed that the number of peak candidates was higher than expected, especially in the {\it HXMT} data. The reason behind this behaviour is the presence of sub-ms spikes caused by the electronics counting repeatedly the large signal deposited by energetic cosmic rays \citep{Wu22}.
Whenever only (i) is fulfilled, we move to a finer time resolution (i.e., 4~ms and then 1~ms if necessary) and look for the same peak until both (i) and (ii) are satisfied; if (i) and (ii) are never satisfied, an upper limit on the FWHM$_{\rm min}$ is taken from the finest timescale for which (i) is satisfied.
When no peak at 64~ms is found that fulfils both (i) and (ii), we move to the $\Delta t=1$~s. If no qualified peak is found, the GRB is discarded because of poor SNR.
Figure~\ref{fig:scheme} shows a schematic description of the procedure.

The FWHM of each peak is estimated through the calibrated relation based on the combination of SNR and  $\Delta t_{\rm det}$ (see Appendix~A for details).
%
%
\begin{table}[]
    \centering
    \begin{tabular}{c|c|c}
    \hline
    Bin time $\Delta t$ (ms)  & SNR$_{\rm min}^{(\Delta t)}$  & $\Delta t_{\rm det,min}^{(\Delta t)}$ (ms)\\
     \hline
    1 & 7.0 & 2\\
    4 & 6.8 & 8 \\
    64 & 6.4 & 128 \\
    1000 & 6.0 & 2000 \\
    \hline
    \end{tabular}
    \caption{Thresholds on SNR and on $\Delta t_{\rm det}$ corresponding to the different bin times, that were adopted to determine the FWHM$_{\rm min}$.}
    \label{tab:threshold}
\end{table}

\section{Results}
\label{sec:res}

\subsection{Comparison between FWHM$_{\rm min}$ and other minimum variability timescale metrics}
\label{subsec:MEPSA_vs_GB14}
\begin{figure}
\centering
\includegraphics[width=9 cm]{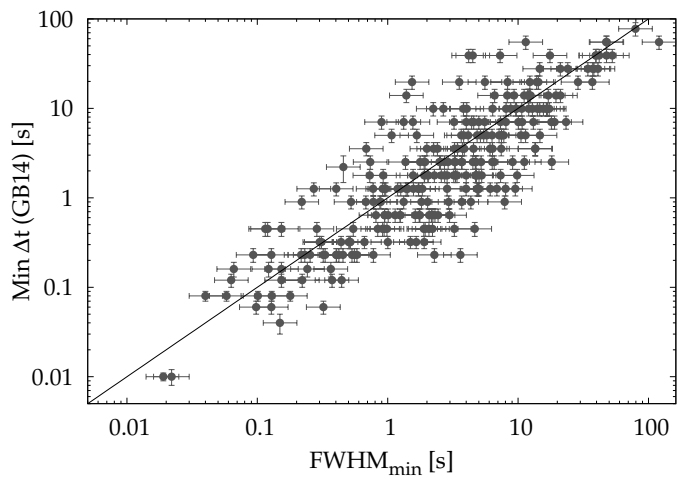}
\caption{Comparison between the shortest FWHM as estimated with {\sc mepsa} (this work) and the MVT estimated by \citetalias{Golkhou14} for a common sample of {\it Swift}/BAT GRBs. Black line corresponds to equality.}
\label{fig:MEPSA_vs_GB14}
\end{figure}

Our choice of adopting FWHM$_{\rm min}$ as an indicator of the MVT of a given GRB LC in principle represents a different definition than other ones that appeared in the literature. In particular, the most distinctive feature of FWHM$_{\rm min}$ is that the duration of a possible, statistically significant, narrow pulse can be enough to qualify as the MVT, irrespective of its impact on the overall variance of the GRB profile and for its physical impact. 
Nevertheless, it is worth exploring how the FWHM$_{\rm min}$ correlates with other metrics. To this aim, we selected a common sample of 467 BAT GRBs, for which \citetalias{Golkhou14} estimated the MVT. Both \citetalias{Golkhou14}'s and our estimates are derived from the same BAT data. Figure~\ref{fig:MEPSA_vs_GB14} shows the comparison between FWHM$_{\rm min}$ and the corresponding MVT estimated by \citetalias{Golkhou14}, with equality shown with a solid line. The two metrics evidently correlate over four decades, with some scatter around equality: this result proves that, although strongly correlated, the two metrics are not completely interchangeable, with a sizeable fraction of cases for which they differ by up to one decade.

%

\subsection{FWHM$_{\rm min}$ as a function of energy}
\label{sec:FWHM_vs_energy}
\begin{figure}[!h]
\centering
\includegraphics[width=9 cm]{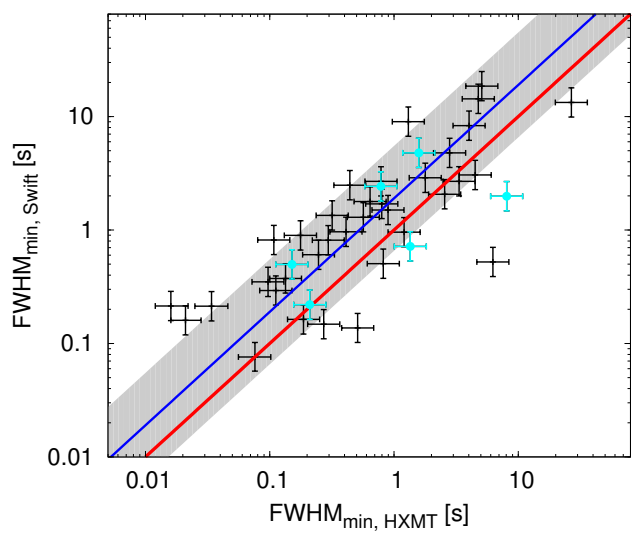}
\caption{FWHM as determined from {\it Insight-HXMT}/HE and {\it Swift}/BAT data for a sample of 44 GRBs in common. The red solid line shows the equality, while the blue line and the shaded area show the best proportionality relation and 1--$\sigma$ region, corresponding to a factor of 2 and a relative scatter of a factor of 3 respectively. Cyan points are GRBs with spectroscopically measured redshift.}
\label{fig:common_BAT_HXMT}
\end{figure}

Figure~\ref{fig:common_BAT_HXMT} shows the comparison of the FWHM$_{\rm min}$ as measured with both {\it Insight-HXMT} and {\it Swift}/BAT using the common sample of 44~GRBs, for three of which only an upper limit on the FWHM$_{\rm min}$ is available. The distribution of the logarithm of the ratio between the two measures is approximately normal, with mean value and standard deviation corresponding to a multiplicative factor of 2 and 3, respectively. Hence, the FWHM$_{\rm min}$ as measured with {\it Swift} is on average twice as long as that measured with {\it Insight-HXMT}. This is similar to the result obtained by \citet{Golkhou15} from the comparison between {\it Fermi}/GBM and {\it Swift}/BAT, as one should expect due to the narrowing of pulses with energy.
%

\begin{figure}
\centering
\includegraphics[width=9 cm]{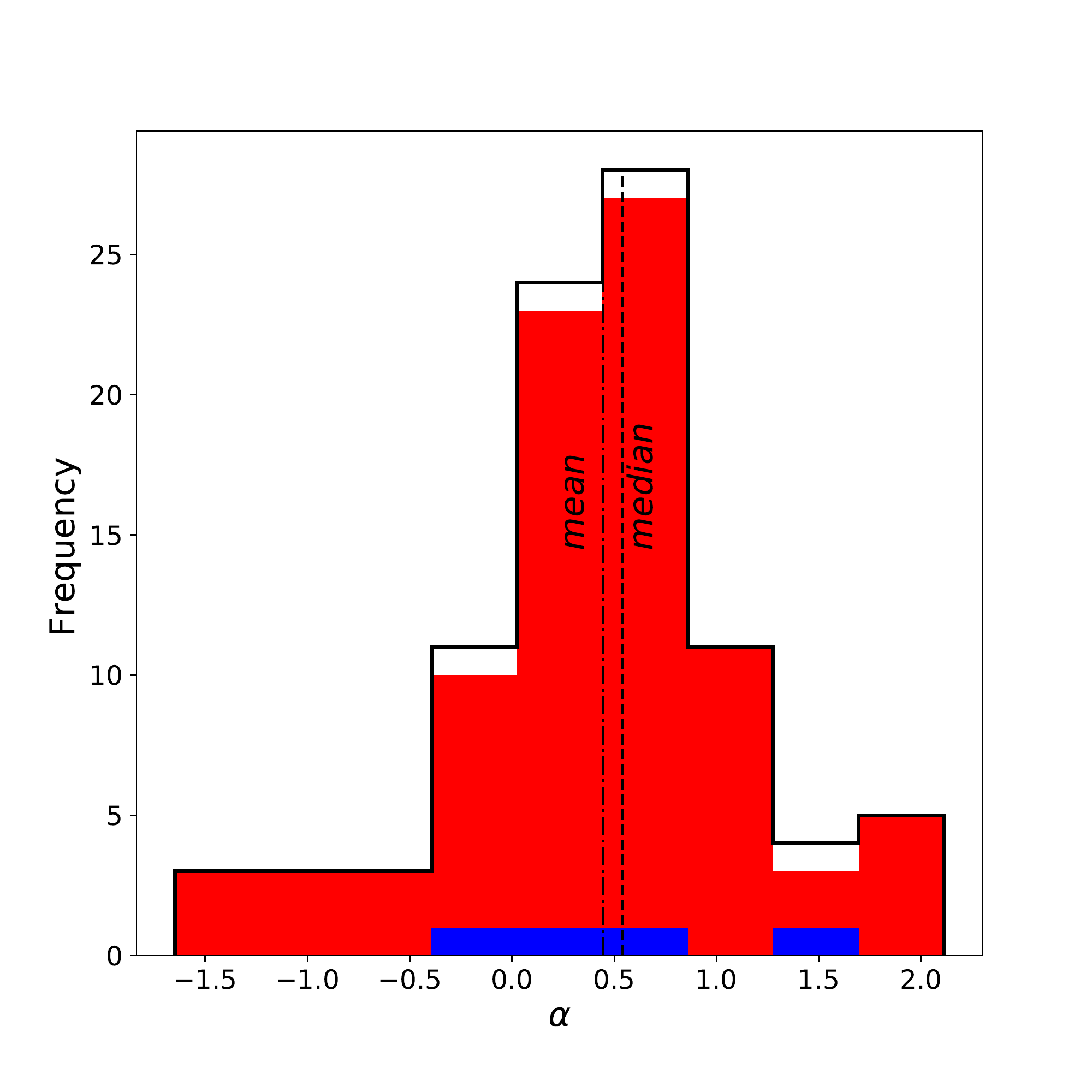}
\caption{Distribution of the power-law index $\alpha$ (FWHM$\,\propto$$E^{-\alpha}$), for all the peaks that were identified in the GRB data of both {\it Swift} and {\it HXMT} (black histogram): SGRBs (filled blue histogram), LGRBs (filled red histogram). Dash-dotted and dashed lines represent the mean ($0.45\pm0.08$) and the median ($0.54\pm0.07$) values, respectively.
}
\label{fig:alpha_distribution}
\end{figure}

Additionally, we can constrain the power-law index $\alpha$ of the relation FWHM$(E) \propto E^{-\alpha}$, where E is the geometric mean of the boundaries of the energy passband. To this aim, for each GRB in the common sample we did not restrict to the shortest pulse of each GRB, instead we calculated $\alpha$ considering the FWHM of the different peaks as detected with {\sc mepsa} in both {\it HXMT} and BAT data. The identification of the same peak, as seen in the two LCs, is assessed through the temporal coincidence within uncertainties.
We identified 93 peaks, 4 of which in SGRBs.
For the energy values, we used the geometric mean of the boundaries of the corresponding  energy bands (15-150 keV for {\it BAT}, 80-800 keV for {\it HXMT} normal mode, 200-3000 keV for {\it HXMT} GRB or low-gain mode;  {\it HXMT} energy bands refer to the deposited energies of incident photons).
We find $\alpha_{\rm mean} = 0.45 \pm 0.08$, $\alpha_{\rm median} = 0.54\pm 0.07$, $\sigma=0.77$: this result is consistent with the values derived modelling the autocorrelation function width \citep{Fenimore95}: $\alpha$ = $0.37$ to $0.43$ (see also \citealt{Borgonovo07}, who found median values in the range $0.21$ to $0.29$).
Figure \ref{fig:alpha_distribution} displays the $\alpha$ distribution for all GRBs (black histogram), SGRBs (blue histogram), LGRBs (red histogram).

\subsection{FWHM$_{\rm min}$ vs. $T_{90}$}
\label{subsec:T90_FWHM}
\begin{figure*}[!h]
\centering
\includegraphics[width=9 cm]{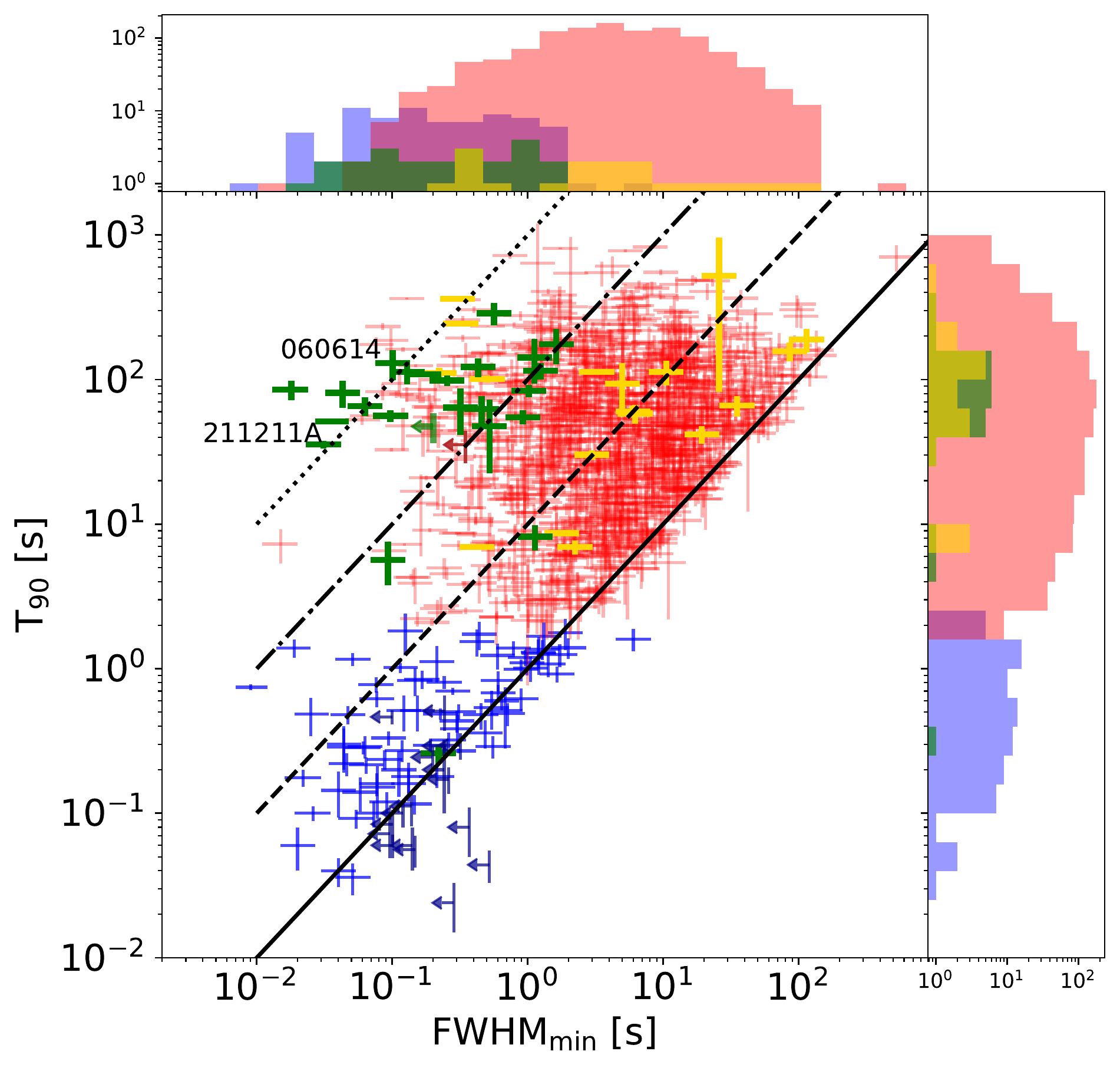}
\includegraphics[width=9 cm]{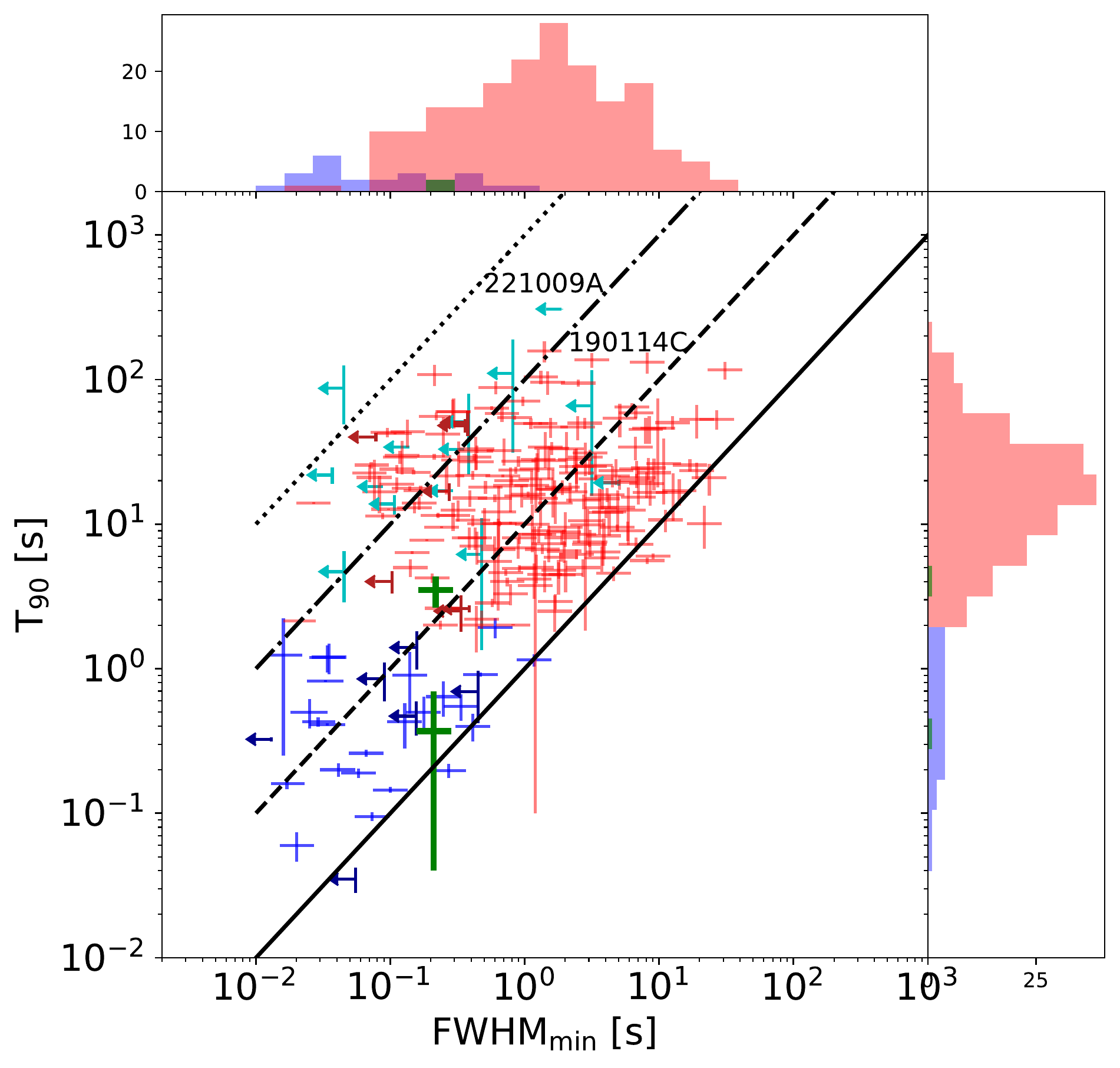}
\caption{Scatter plot of FWHM$_{\rm min}$ and $T_{90}$ for the {\it Swift}/BAT sample (left panel) and for the {\it Insight-HXMT} sample (right panel) along with the corresponding marginal distributions for three distinct populations of SGRBs (blue), LGRBs (red), SEE-GRBs (green). Gold points are LGRBs with an associated SN. We highlight two SEE-GRBs, 060614 and 211211A, for which there is strong evidence for a binary merger rather than a collapsar origin. Cyan points refer to the {\it Insight-HXMT} saturated GRBs. We highlight 190114C and 221009A because of their high extreme energy \citep{MAGIC19a,Frederiks22}. We do not consider {\it Swift} data where $T_{90}< \sigma_{T_{90}}$. Solid, dashed, dash-dotted, and dotted lines represent the equality, $10^{1}$, $10^{2}$, and $10^{3}$ factor, respectively.}
\label{fig:FWHM_T90_HXMT}
\end{figure*}

Figure \ref{fig:FWHM_T90_HXMT} shows the scatter plots of FWHM$_{\rm min}$ and $T_{90}$ for {\it Swift}/BAT and {\it Insight-HXMT} samples along with the corresponding marginal distributions. FWHM$_{\rm min}$ spans the range from $10^{-2}$ to $10^2$~s, whereas $T_{90}$ from $10^{-2}$ to $10^3$~s ($10^2$~s for {\it Insight-HXMT}). 

The bimodal nature of the marginal distribution of $T_{90}$ seems to be slightly more evident in the {\it Insight-HXMT} sample, in spite of the fewer GRBs.
Equality is shown with a solid line. In addition, to guide the eye, we show lines that mark constant values for the ratio $r = T_{90}/$FWHM$_{\rm min}$: $10$ (dashed), $100$ (dash-dotted), $1000$ (dotted). Clearly, $r$ increases with an increasing number of pulses within a GRB and/or with the presence of quiescent times. Most single-pulse GRBs lie in the region $1\lesssim r< 10$. Unsurprisingly, almost all SGRBs lie within this region. LGRBs instead span the range $1\le r\lesssim 10^3$.

The marginal distributions of FWHM$_{\rm min}$ of SGRBs and of LGRBs are evidently different: for the {\it Swift} sample a Kolmogorov-Smirnov (KS) test\footnote{The two-sample KS test was done using {\tt scipy.stats.ks\_2samp}.} yields a probability of $10^{-37}$ of being drawn from the same population: the logarithmic means of SGRBs and of LGRBs are respectively $0.2$ and $4$~s with a comparable scatter of $0.6$~dex.
Also in the {\it Insight-HXMT} sample the two classes of S- and LGRBs have significantly different FWHM$_{\rm min}$ distributions, with a probability of $3\times 10^{-12}$ of being drawn from a common population. The logarithmic mean values are $0.1$ and $1.3$~s for the S- and LGRBs with the same scatter of $\sim 0.6$~dex, respectively, i.e. shorter than the corresponding quantities obtained in the softer energy band of the {\it Swift} sample, in line with the results of Section~\ref{sec:FWHM_vs_energy}.

In the {\it Swift} sample we highlight the population of SEE-GRBs (green), with emphasis on two peculiar events, whose LC and duration look like a typical LGRB one, but for which robust evidence for a compact binary merger progenitor rather than a collapsar was found: 060614 \citep{Gehrels06,DellaValle06,Fynbo06,Jin15} and 211211A \citep{Rastinejad22,Gompertz22,Yang22,Troja22}. In spite of being just 24, their FWHM$_{\rm min}$ values are more similar to those of SGRBs than of LGRBs. While both KS and Anderson-Darling (AD) \footnote{The two-sample AD test was done using {\tt scipy.stats.anderson\_ksamp}.} tests between the FWHM$_{\rm min}$ of SGRBs and SEE-GRBs do not reject the common population null hypothesis, the comparison between LGRBs and SEE-GRBs does with $3.4\times 10^{-14}$ KS-probability (AD probability $<10^{-3})$. Therefore, the FWHM$_{\rm min}$ is a promising metric to identify SEE-GRBs, even when the LC does not look like a single spike followed by a long, soft and smooth tail, but rather shows multiple peaks extending over several seconds, as was the case of 060614 and 211211A.
This suggests that the contamination of long GRBs having FWHM$_{\rm min}\lesssim0.1$~s that are currently misclassified as Type-II GRBs, could be higher than expected, as was recently put forward from the observations of events like 211211A.

As a further check, we also show 18 LGRBs for which the collapsar origin is not disputable, thanks to the identification of an associated SN (gold points).\footnote{They are: 060729, 090618, 091127, 100316D, 101219B, 111228A, 120422A, 120714B, 120729A, 130215A, 130427A, 130831A, 140506A, 161219B, 171205A, 180728A, 190114C, and 190829A.} Although a few of them have a FWHM$_{\rm min}$ which is comparable with that of SEE-GRBs, the bulk of SN-associated GRBs have longer MVT, more representatively of the entire population of Type-II GRBs. Furthermore, the comparison between LGRBs associated with SN and SEE-GRBs done with KS and AD tests rejects the common population null hypothesis with a probability of $1.2\times 10^{-5}$ ($<10^{-3}$) for KS (AD).

In {\it Insight-HXMT} plot  (see Figure \ref{fig:FWHM_T90_HXMT}, right panel) we added the saturated GRBs (cyan), highlighting the cases of 190114C and 221009A; for 5\footnote{They are: 181222B, 190606A, 191025B, 191227B, 201227A.} of the 21 saturated GRBs it was not possible to find any eligible peak: these GRBs are SGRB and the time windows which are not saturated are too short for estimating FWHM$_{\rm min}$.

\subsection{Dependence of FWHM$_{\rm min}$ and $T_{90}$ on redshift}
\label{sec:FWHM_vs_z}
For the BAT sample with measured redshift, we studied whether the $T_{90}$ and FWHM$_{\rm min}$ distributions show evidence for cosmological time dilation. We calculated the geometric mean of groups of GRBs: 30 (6) GRBs each for LGRBs (SGRBs) and then fitted the geometric mean values as a function of $(1+z)^\alpha$, with the power-law index $\alpha$ free to vary.
\begin{figure*}[h]
\centering
\includegraphics[width=9 cm]{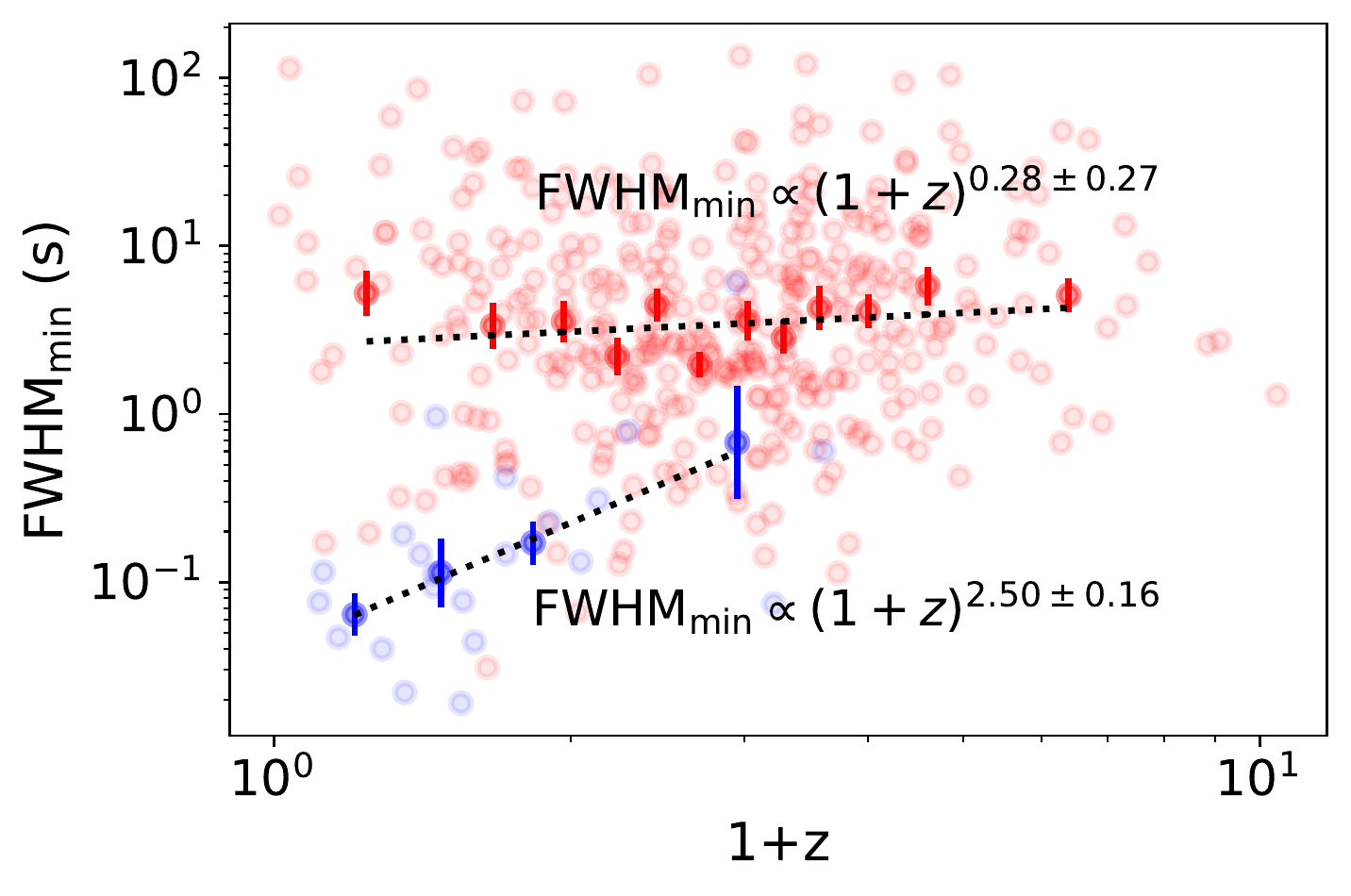}
\includegraphics[width=9 cm]{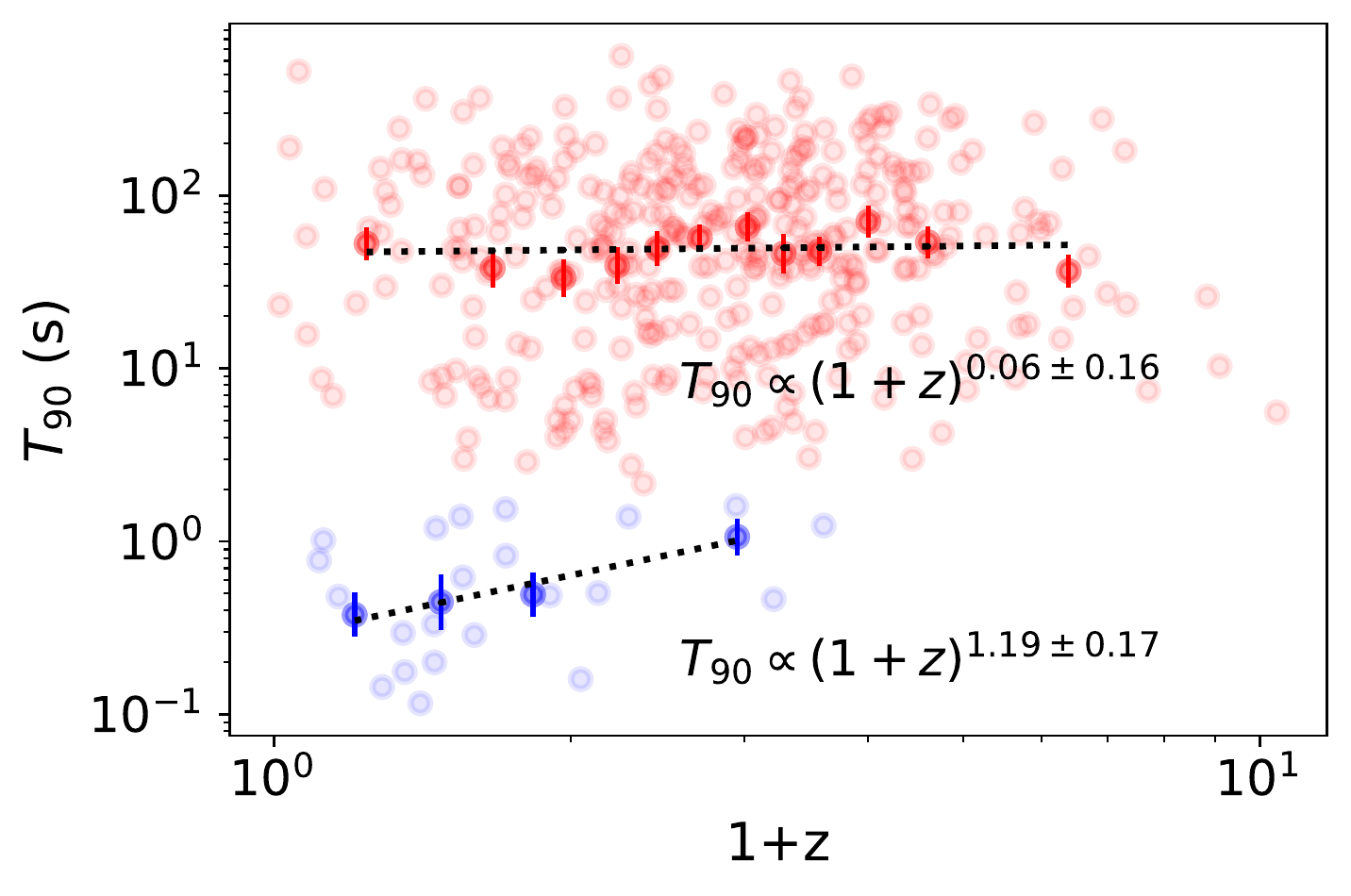}
\caption{FWHM$_{\rm min}$ vs. $(1+z)$ (left panel) and $T_{90}$ vs. $(1+z)$ (right panel) for the {\it Swift} sample as seen in the observer frame. 
Red (blue) dots correspond to LGRBs (SGRBs). Lighter dots refer to individual GRBs; darker ones refer to $T_{90}$ geometrical mean of groups of 30 (6) LGRBs (SGRBs), sorted with increasing $z$.}
\label{fig:FWHM_T90_z}
\end{figure*}
Figure~\ref{fig:FWHM_T90_z} shows the results, which suggest that there is no evidence for a dependence on redshift ($\alpha$ being compatible with zero), at least as long LGRBs are concerned. The reason is that the different effects at play combine in such a way that the cosmological dilation is not dominant, as also found by previous investigations (\citealt{KocevskiPetrosian03}; \citetalias{Golkhou14}; \citealt{Golkhou15}; \citealt{Littlejohns14b}). In particular, the effects that contribute to mask the impact of cosmic dilation are mainly two: (a) for a given observer energy passband, further GRBs are observed in a rest-frame harder band, which implies narrower temporal structures, according to the narrowing of pulses with energy \citep{Norris96,Fenimore95}; (b) the SNR of a LC decreases with increasing $z$ so that only the brightest portion of the LC can be detected \citep{KocevskiPetrosian03}.

For SGRBs the results seem to indicate values of both FWHM$_{\rm min}$ and $T_{90}$ increasing with redshift. The reason for this apparently different behaviour from LGRBs is not obvious; in particular it is not clear why the same effects mentioned above for LGRBs do not appear to affect SGRBs in a comparable way. It is nonetheless worth noting that each SGRB group includes just 6 events, a choice forced by the small number of SGRBs with known $z$, so our accuracy in estimating the standard deviation of FWHM$_{\rm min}$ and $T_{90}$ for SGRB groups is lower than for LGRBs.
In any case, the redshift range is shorter than that of LGRBs, so in principle the cosmic dilation correction does not have the same impact as it should for LGRBs, notwithstanding that there are counteracting effects at play. Moreover, should we correct the FWHM$_{\rm min}$ values for the SGRB sample, the result of statistically different distributions of FWHM$_{\rm min}$ for LGRBs and SGRBs would be reinforced.
For this reason, we conservatively avoided correcting the observer-frame value for FWHM$_{\rm min}$ for cosmological time dilation.

\subsection{Peak rate vs. FWHM$_{\rm min}$}
\label{sec:PR_vs_FWHM}

\begin{figure*}[tb]
\centering
\includegraphics[height=7.3 cm]{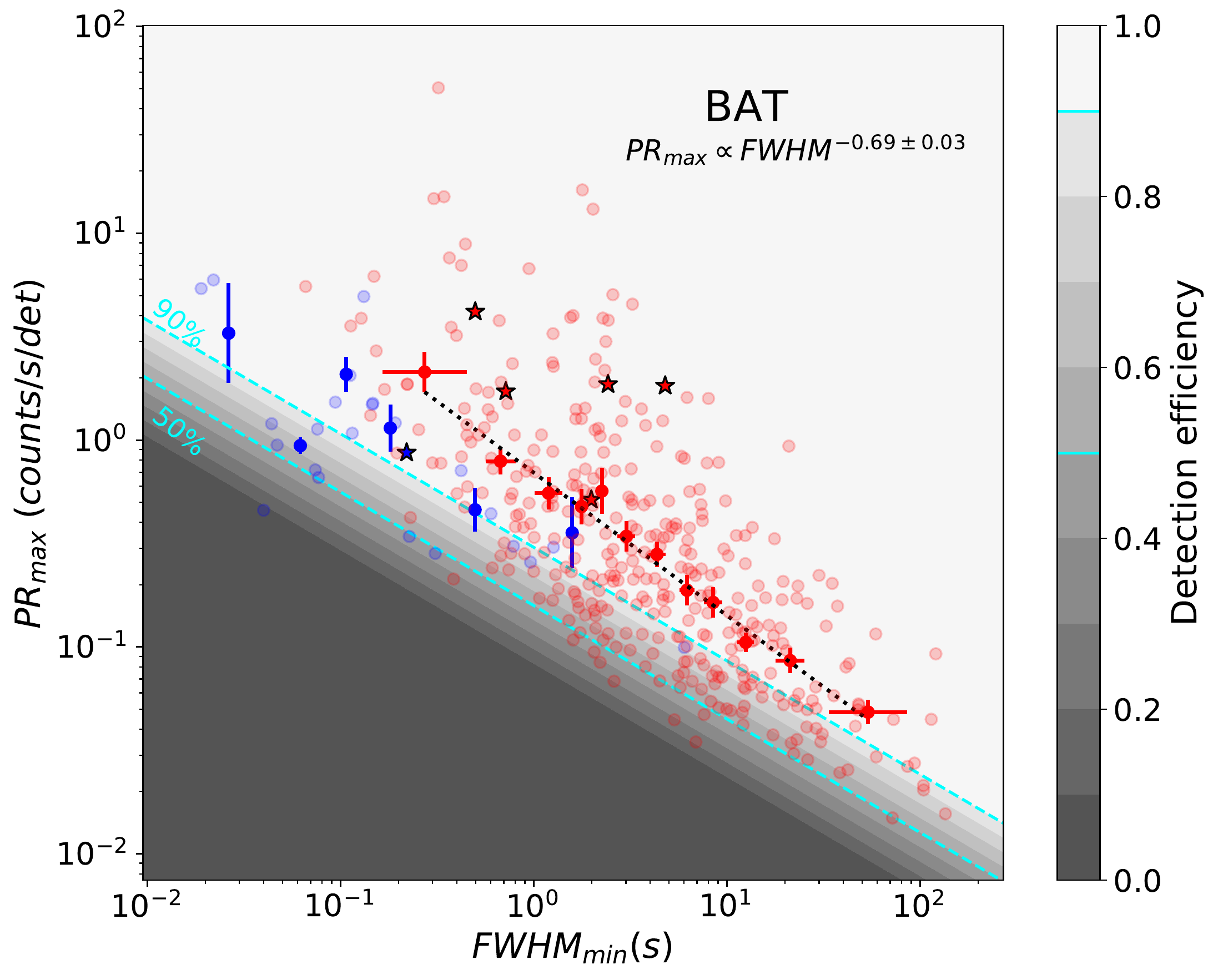}
\includegraphics[height=7.3 cm]{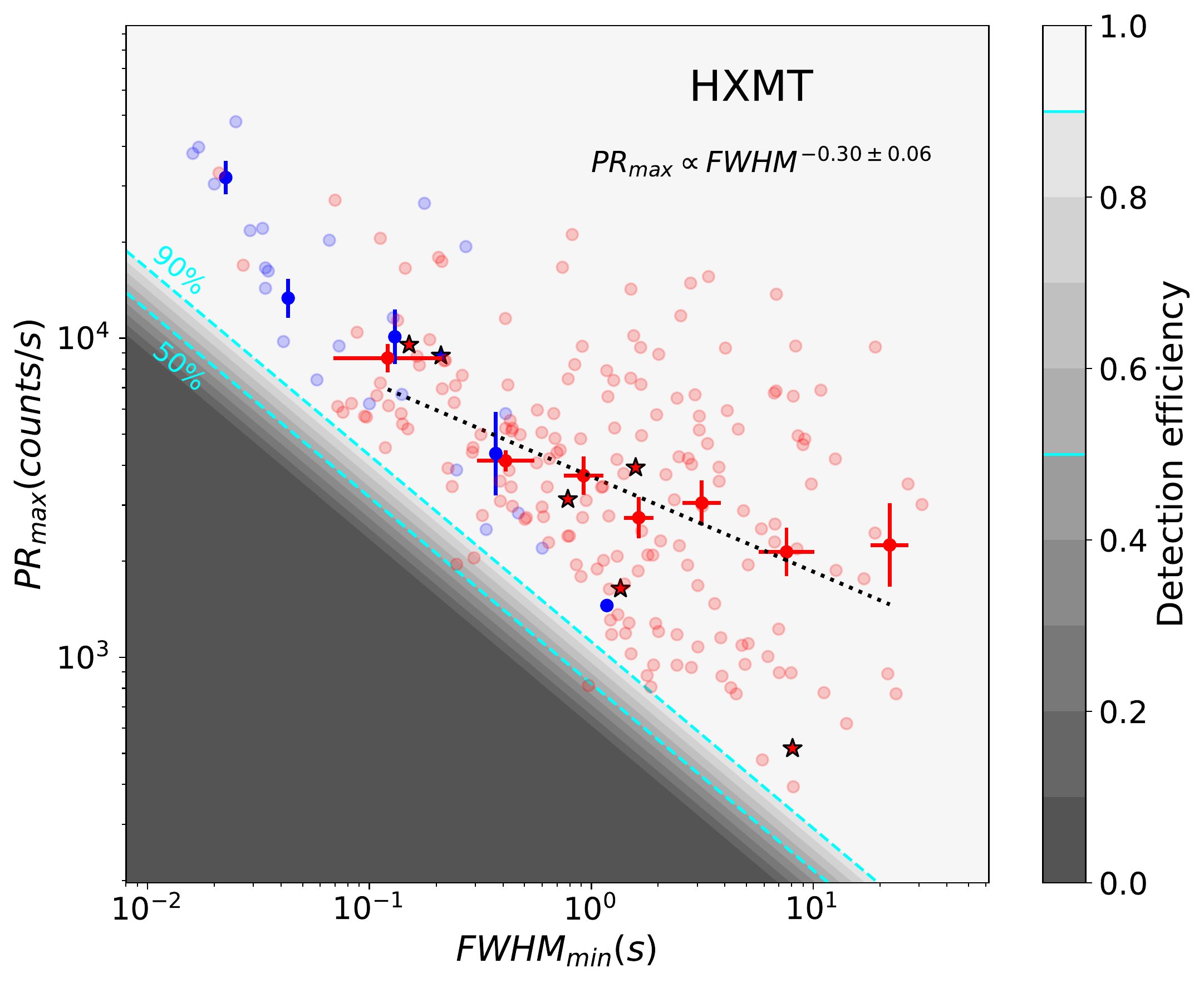}
\caption{Peak rate vs. FWHM$_{\rm min}$ for the {\it Swift} (left) and {\it HXMT} (right) samples. Blue dots correspond to Type-I GRBs (i.e., SGRBs and SEE-GRBs), red ones to Type-II GRBs. Lighter dots refer to single GRB data, darker ones refer to geometrical mean of data of groups of GRBs sorted with increasing z: for Type-I each group is composed by 6 GRBs, for Type-II each group is composed by 30 GRBs. Stars refer to the 6 GRBs in common between {\it Swift} sample and {\it HXMT} sample. Dotted lines indicate the best fit of type II GRBs. The shaded areas show ten different regions with detection efficiency spanning the 0 to 1 range. Cyan dashed lines show the 50\% and 90\% detection efficiency contours.}
\label{fig:PR_FWHM}
\end{figure*}

The peak rate PR$_{\rm max}$ is defined as the peak rate of the most intense pulse of a given GRB. To explore the relation between isotropic-equivalent peak luminosity ($L_{\rm p}$) and FWHM$_{\rm min}$ we first need to understand the relation between PR$_{\rm max}$ and FWHM$_{\rm min}$ focusing on the selection effects that inevitably come into play. We do not consider the cosmological time dilation in its simplest formulation, which is FWHM$_{\rm min}/(1+z)$, given the interplay of different effects already discussed in Sect.~\ref{sec:FWHM_vs_z}.  
PR$_{\rm max}$ appears to be anticorrelated with FWHM$_{\rm min}$ (Fig.~\ref{fig:PR_FWHM}). However, the selection effect here at play is to be evaluated carefully: for a very short pulse to be detected, its peak rate must be high enough to ensure the required minimum SNR. To this aim, we simulated a number of Fast Rise Exponential Decay (FRED) pulses covering the interested PR$_{\rm max}$ vs. FWHM$_{\rm min}$ space and counted the fraction of pulses that were identified by our procedure. As a result, we modelled the  detection efficiency $\epsilon_{\rm det}$, defined as the fraction of simulated pulses that are correctly identified, as a function of both PR$_{\rm max}$ and FWHM$_{\rm min}$: $\epsilon_{\rm det}({\rm PR}_{\rm max}, {\rm FWHM}_{\rm min})$. We found that the dependence is approximately linear in the logarithms for both {\it Swift} and {\it HXMT} data sets:
\begin{equation}
 \epsilon_{\rm det}\ =\ a\,\log_{10}{\left(\frac{{\rm FWHM}_{\rm min}}{\rm s}\right)}\ +\ b\,\log_{10}{\left(\frac{{\rm PR}_{\rm max}}{\rm rate\ unit}\right)}\ +\ c\;,
\label{eq:deteffBAT}
\end{equation}
where the rate units are either counts~s$^{-1}$~det$^{-1}$ ({\it Swift}/BAT) or counts~s$^{-1}$ ({\it HXMT}/HE). For BAT, the best-fit parameters are $a=0.78$, $b=1.42$, and $c=1.64$, and it is clipped either to 0 for negative values, or to 1 for values exceeding it.
One can conveniently invert Equation~(\ref{eq:deteffBAT}) to express the minimum value for PR$_{\rm max}$ required for a pulse with a given FWHM$_{\rm min}$ to be detected, which in the case of {\it Swift}/BAT becomes
\begin{equation}
 {\rm PR}_{\rm max}^{\rm (BAT)}\ \ge\ 0.35\ {\rm cts/s/det}\ \ \ {\left(\frac{{\rm FWHM}_{\rm min}}{\rm s}\right)}^{-0.55}\ 10^{0.7\,(\epsilon-1)}\;.
\label{eq:deteffBAT_inv}
\end{equation}

Figure~\ref{fig:PR_FWHM} shows the results for both data sets: {\it Swift}/BAT (left panel) and {\it HXMT} (right panel) as colour-coded shaded area with detection efficiency being split into ten different ranges, from 0 to 1. This selection effect clearly affects the observed correlation between the two quantities for both classes of SGRBs and LGRBs.
%

Replicating similar simulations and analysis for the {\it HXMT} sample, we obtained a similar result which is described by Eq.~(\ref{eq:deteffBAT}) with the following parameters: $a=1.78$, $b=3.05$, and $c=-8.40$. The analogous condition of Eq.~(\ref{eq:deteffBAT_inv}) for {\it HXMT} is therefore
\begin{equation}
 {\rm PR}_{\rm max}^{\rm (HXMT)}\ \ge\ 1200\ {\rm cts/s}\ \ \ {\left(\frac{{\rm FWHM}_{\rm min}}{\rm s}\right)}^{-0.58}\ 10^{0.33\,(\epsilon-1)}\;.
\label{eq:deteffHXMT_inv}
\end{equation}
%

\subsection{Peak luminosity vs. FWHM$_{\rm min}$}
\label{sec:Lp_vs_FWHM}
When we move from the observer to the GRB rest frame by replacing PR$_{\rm max}$ with the intrinsic quantity $L_p$, the problem of accounting for this bias becomes more complicated, due to the wide range in luminosity distance $D_L$ spanned by the subsample with measured redshift. In fact, a given peak rate corresponds to a range in luminosities which is dominated by the range spanned by $D_L^2$.

Given the paucity of GRBs with measured redshift in the {\it HXMT} sample, we analysed $L_{\rm p}$ as a function of FWHM$_{\rm min}$ only for the {\it Swift} sample with known $z$. 
We estimated $L_{\rm p}$ for this sample by renormalising the peak count rate by the ratio between total counts and the fluence  (in units of erg~cm$^{-2}$ in the 15-150~keV energy band), as published by the BAT team.\footnote{\url{https://swift.gsfc.nasa.gov/archive/grb_table/}.} This procedure assumes a time-averaged spectrum, which is equivalent to neglecting any spectral evolution. However, for our purposes, the impact of an error on $L_{\rm p}$ up to a factor of a few is overall negligible. 

\begin{figure*}
\centering
\includegraphics[width=18cm]{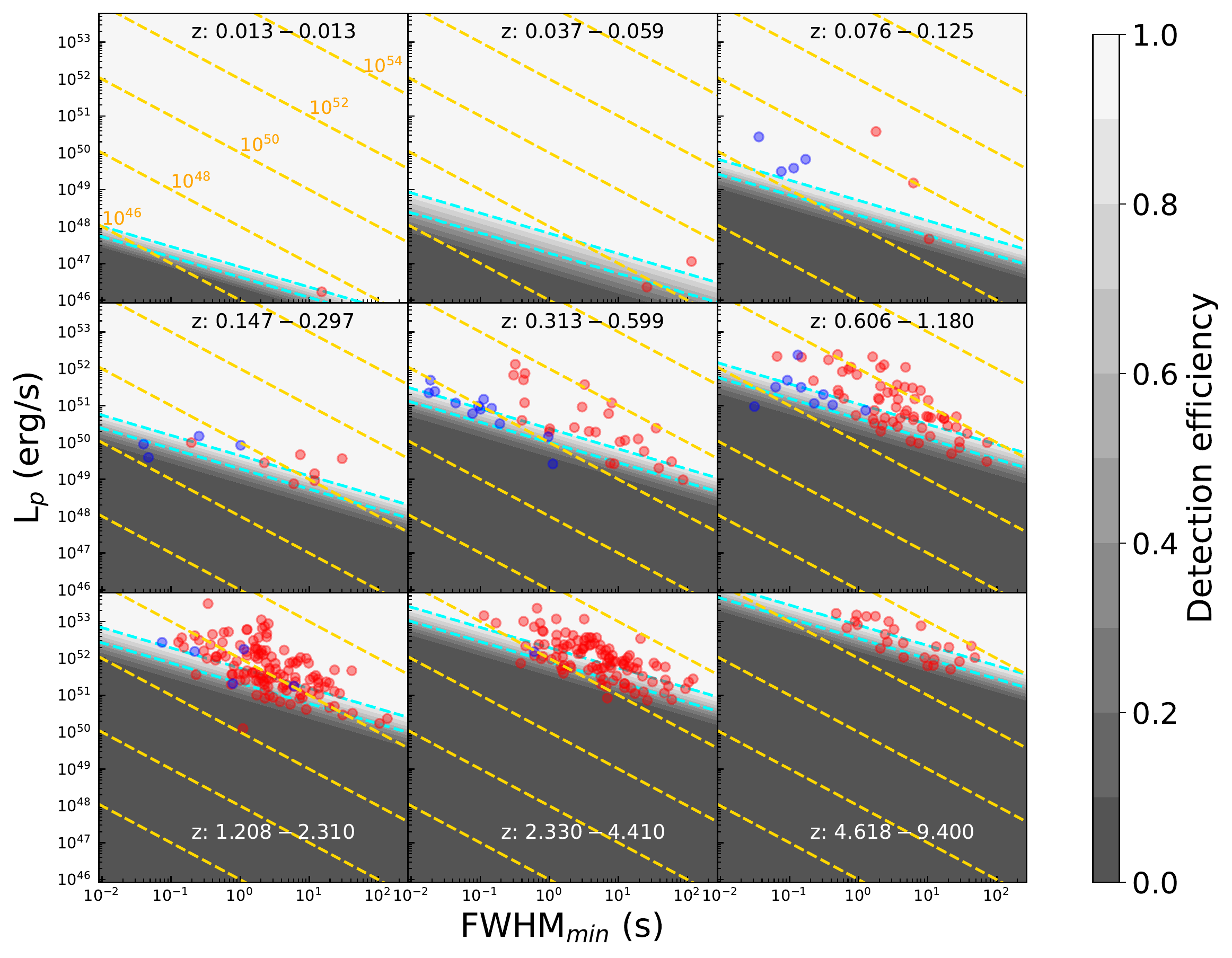}
\caption{Peak Luminosity vs. FWHM$_{\rm min}$ for the {\it Swift}/BAT sample split into 9 redshift bins with constant logarithmic space in luminosity distance; blue dots correspond to Type-I GRBs, red ones to Type-II GRBs. Dashed cyan lines correspond to 90\% and 50\% detection efficiency (vertical bar). Gold dashed lines represent loci of constant isotropic-equivalent released energy (in erg) of each individual peak, approximately calculated as $E_{\rm iso}=L_p$~FWHM$_{\rm min}$.} 
\label{fig:Lp_FWHM_BAT}
\end{figure*}

We split the sample into 9 subsets with luminosity distance evenly spaced logarithmically and considered the impact of detection efficiency in the $L_p$--FWHM$_{\rm min}$ plane independently of each other. The results are shown in Figure.~\ref{fig:Lp_FWHM_BAT}: the selection effect clearly affects the final correlation. In particular, two properties are evident: (i) the absence of relatively low luminosity GRBs with short FWHM$_{\rm min}$ (bottom left corner of each panel) is clearly an observational bias; (ii) the absence of luminous GRBs with long FWHM$_{\rm min}$ (top right corner of each panel) is an intrinsic propriety of GRBs and suggests the possible existence of a maximum radiated energy within a single pulse (e.g. \citealt{Dado22}). To better appreciate this possibility, in each panel we show a grid with constant isotropic-equivalent released energy values of a single pulse, roughly estimated as $E_{\rm iso}^{\rm (pulse)}\approx L_p$~FWHM$_{\rm min}$.

Taken at face value, for the sample of BAT Type-I GRBs with measured $z$ there is no evidence for correlation between $L_p$ and FWHM$_{\rm min}$, possibly due to the relatively few GRBs. On the contrary, in the case of Type-II GRBs the two observables do correlate: null-hypothesis probabilities given by Pearson's linear (calculated on logarithmic values), Spearman's and Kendall's rank correlation tests are $1.5\times10^{-21}$, $2.6\times10^{-23}$, and $9.8\times10^{-22}$, respectively. However, the key aspect is understanding the extent by which this correlation is driven by the selection bias.

\begin{figure}
\centering
\includegraphics[width=9cm]{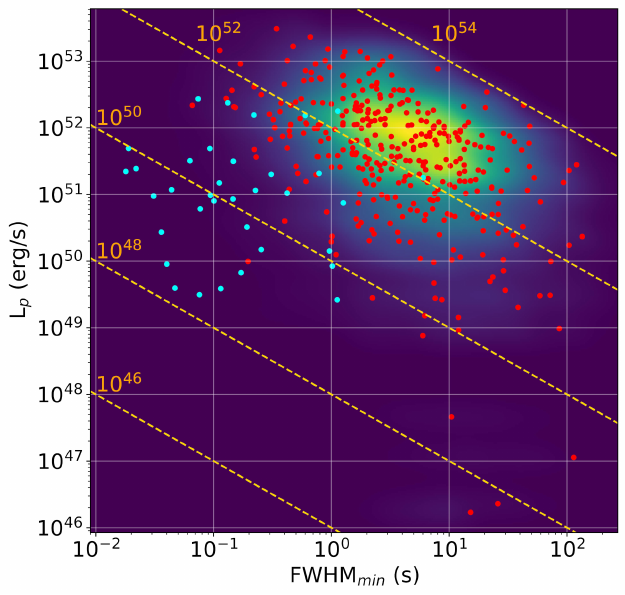}
\caption{Peak Luminosity $L_p$ vs. FWHM$_{\rm min}$ for the {\it Swift}/BAT Type-II GRB sample (red circles). The colour-coded density map is the result of a simulated sample of Type-II GRBs that accounts for the selection effects shown in Figure~\ref{fig:Lp_FWHM_BAT} under the assumption of no correlation between $L_p$ and FWHM$_{\rm min}$. For comparison, we also show SGRBs (cyan circles). Gold dashed lines represent loci with constant isotropic-equivalent released energy of each individual peak, approximately calculated as $E_{\rm iso}=L_p$~FWHM$_{\rm min}$.} 
\label{fig:Lp_FWHM_BAT_LGRBs_sim}
\end{figure}

To this aim, we carried out a suite of simulations under the assumption that $L_p$ is distributed independently of FWHM$_{\rm min}$ and hence there exists no intrisic correlation at all. For each of the 9 subsets with different redshift bins we randomly generated as many points in the FWHM$_{\rm min}$-$L_p$ plane as those in the true subset, taking the very same values of $L_p$ and assuming for FWHM$_{\rm min}$ a Gaussian kernel density estimation\footnote{We used the {\tt scipy.stats.kde.gaussian\_kde()} function.} of the observed FWHM$_{\rm min}$ distribution of the entire LGRB sample with measured $z$ as the probability density distribution. For each point to be accepted, we further set two conditions: 1) we randomly ran a Bernoulli trial assuming a probability given by the detection efficiency calculated at that point according to Eq.~(\ref{eq:deteffBAT}) and only the points with an outcome of 1 were taken; 2) $L_p$~FWHM$_{\rm min} \le E_{\rm iso,max}^{\rm (pulse)}$, i.e. we demanded that the isotropic-equivalent pulse energy should not overcome the highest value observed in the corresponding real subsample. For each subset trials went on until the accepted simulated points were as many as the real ones. The rationale of 1) is mimicking the effect of the selection bias, whereas the purpose of 2) is accounting for the absence of very luminous and long FWHM$_{\rm min}$ that is observed in the real sample.
Every such round ended up with a simulated sample that shared the same number of Type-II GRBs as the real one and affected by the same selection bias, with no intrinsic correlation between $L_p$ and FWHM$_{\rm min}$. We simulated $10^4$ such samples and derived a density map in the FWHM$_{\rm min}$--$L_p$ plane to be compared with the real sample.
Figure~\ref{fig:Lp_FWHM_BAT_LGRBs_sim} shows the result. While the synthetic population does exhibit a correlation, as expected, it also appears more clustered than the real set. 
%

\begin{figure}
\centering
\includegraphics[width=9cm]{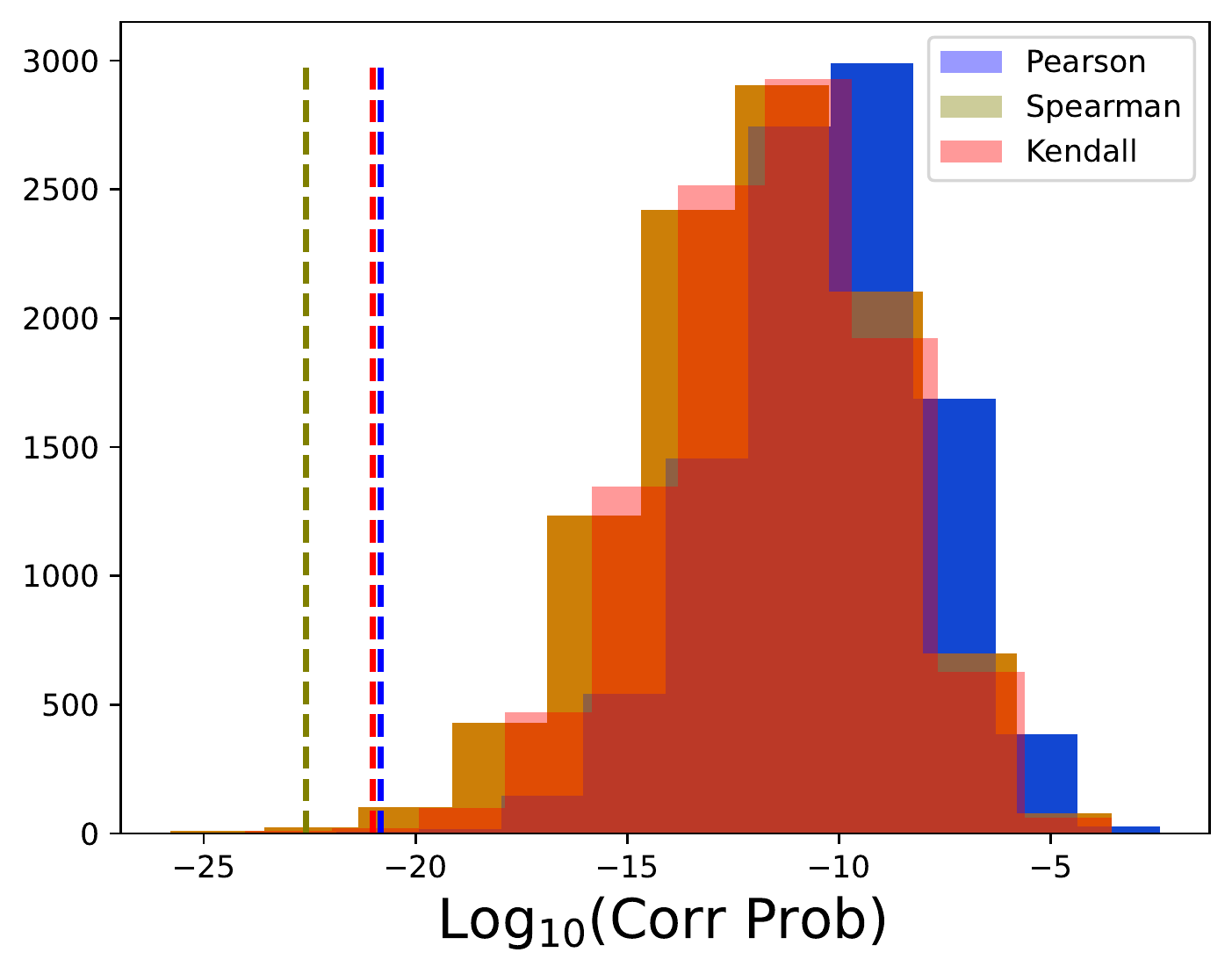}
\caption{FWHM$_{\rm min}$--$L_p$ correlation probability distributions of the simulated samples of Type-II GRBs, compared with the corresponding values obtained for the real sample of Type-II GRBs with measured $z$ (vertical lines).} 
\label{fig:Lp_FWHM_BAT_LGRBs_sim_corrcoeff}
\end{figure}

To quantitatively assess how compatible the real and all the simulated samples are, we adopted the following approach: for each of them we calculated the three correlation coefficients as we did for the real sample and obtained the corresponding distributions. Figure~\ref{fig:Lp_FWHM_BAT_LGRBs_sim_corrcoeff} shows the p-value distributions of the simulated samples, compared with the analogous quantities obtained for the real sample.
The real sample exhibits a more significant correlation than what simulated samples do: for each of the three correlation tests, only $\le 20$~synthetic samples out of $10^4$ exhibited lower p-values than the real sample.

%

We further explored how tightly this result depends on the assumed distribution of FWHM$_{\rm min}$: in fact, the observed one is inevitably the result of the same selection bias, whose impact we aim to study. We then repeated the same suite of simulations by assuming broader FWHM$_{\rm min}$ distributions, in particular extending towards small values. As a result, the comparison between the simulated and the real correlation coefficients did not change essentially, so the conclusion that the observed correlation cannot be entirely ascribed to the selection bias still holds. 

\subsection{Number of peaks vs. FWHM$_{\rm min}$}
\label{sec:Npeaks_vs_FWHM}
\begin{figure*}
\centering
\includegraphics[width=9 cm]{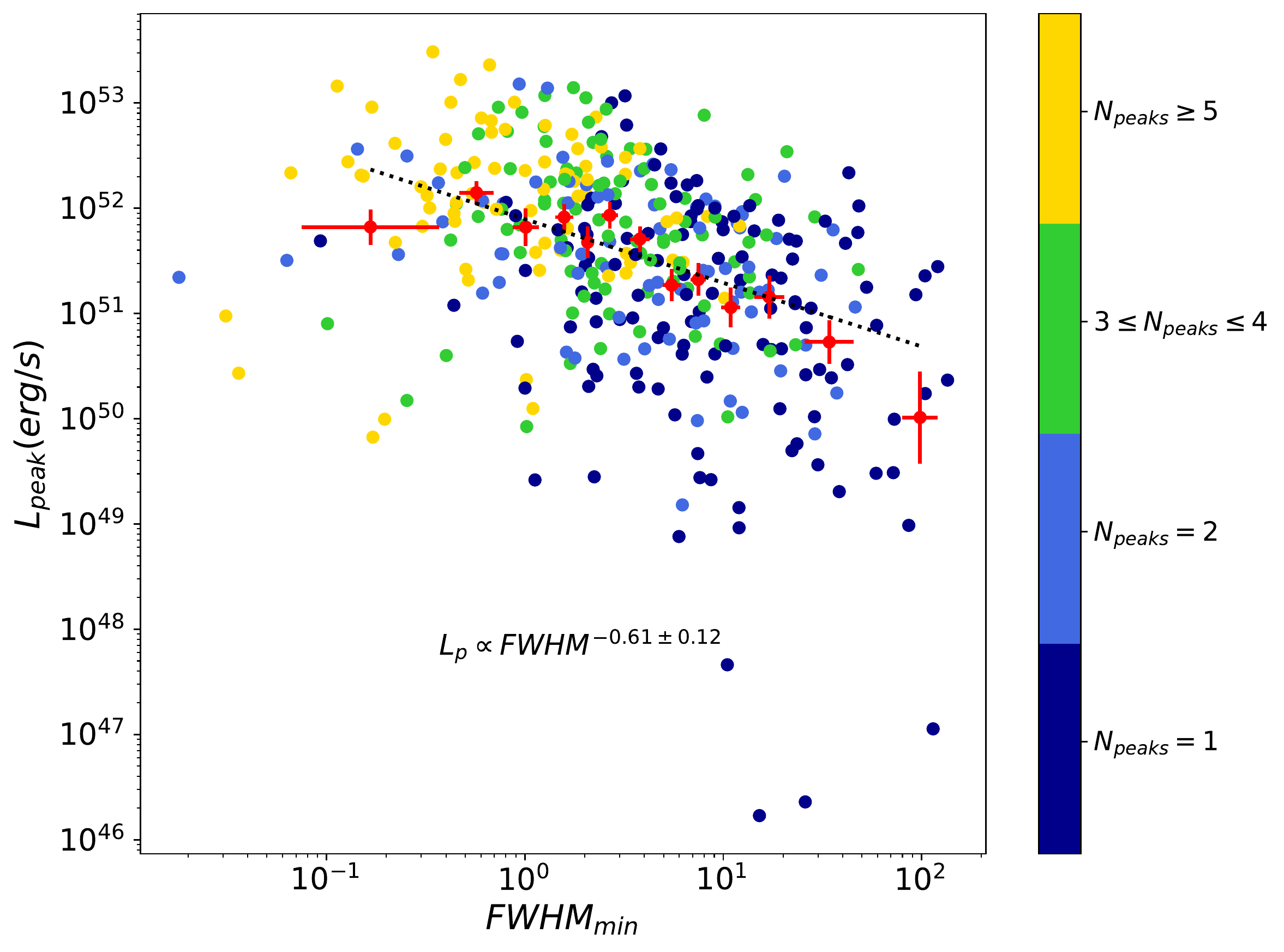}
\includegraphics[width=9 cm]{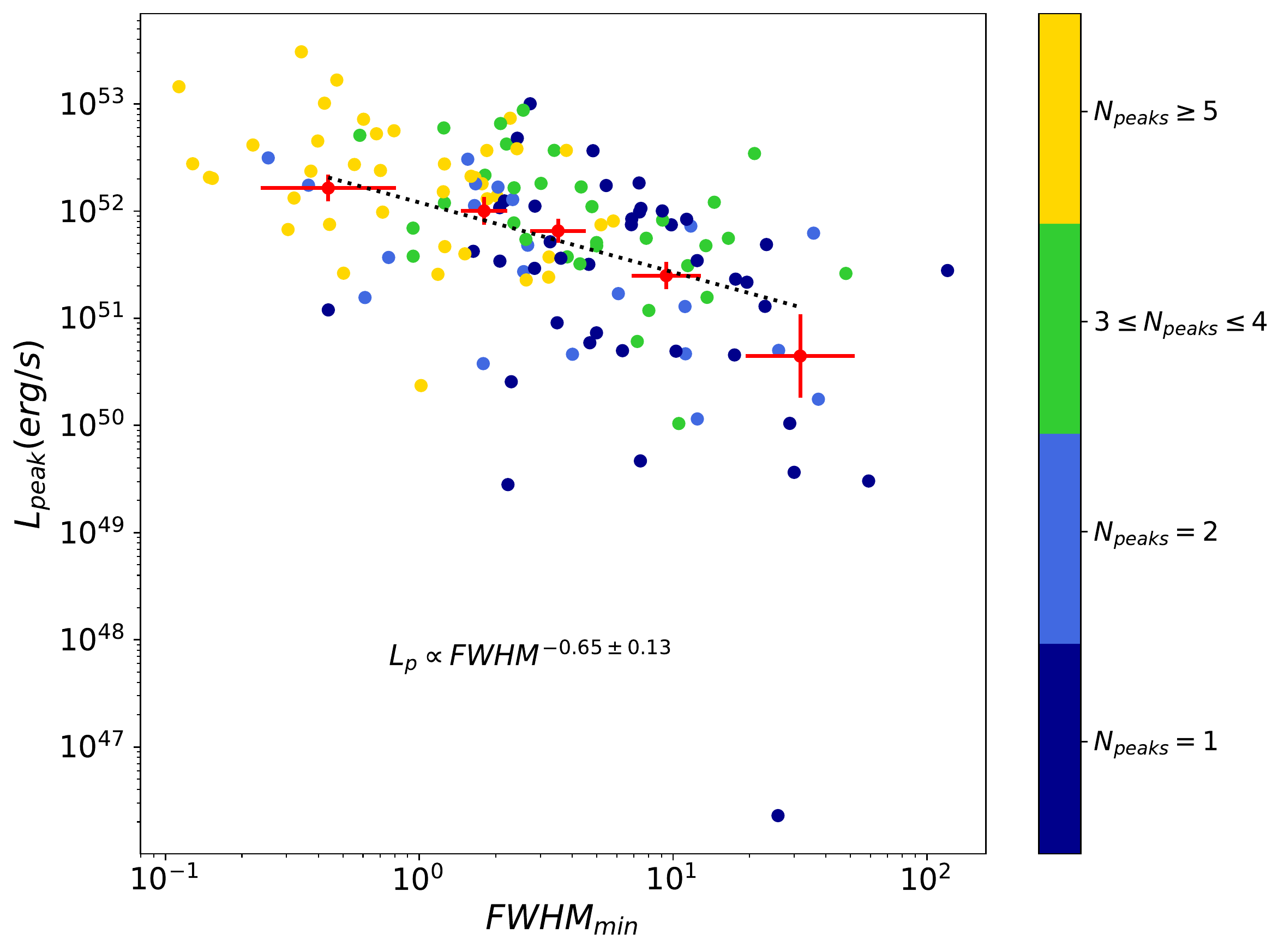}
\caption{Peak Luminosity vs. FWHM$_{\rm min}$ for the {\it Swift} sample of Type-II GRBs for different number of peaks $N_{peaks}$ (left panel); $L_{peak}$ vs. FWHM$_{\rm min}$ for the {\it Swift} sample of Type-II GRBs, SNR > 12.26 (right panel), for different number of peaks $N_{peaks}$. Red dots refer to geometrical mean of groups of GRBs, with each group including 30 GRBs, independently on $N_{peak}$.}
\label{fig:Lp_FWHM_differentNpeaks}
\end{figure*}

\begin{figure}
\centering
\includegraphics[width=9 cm]{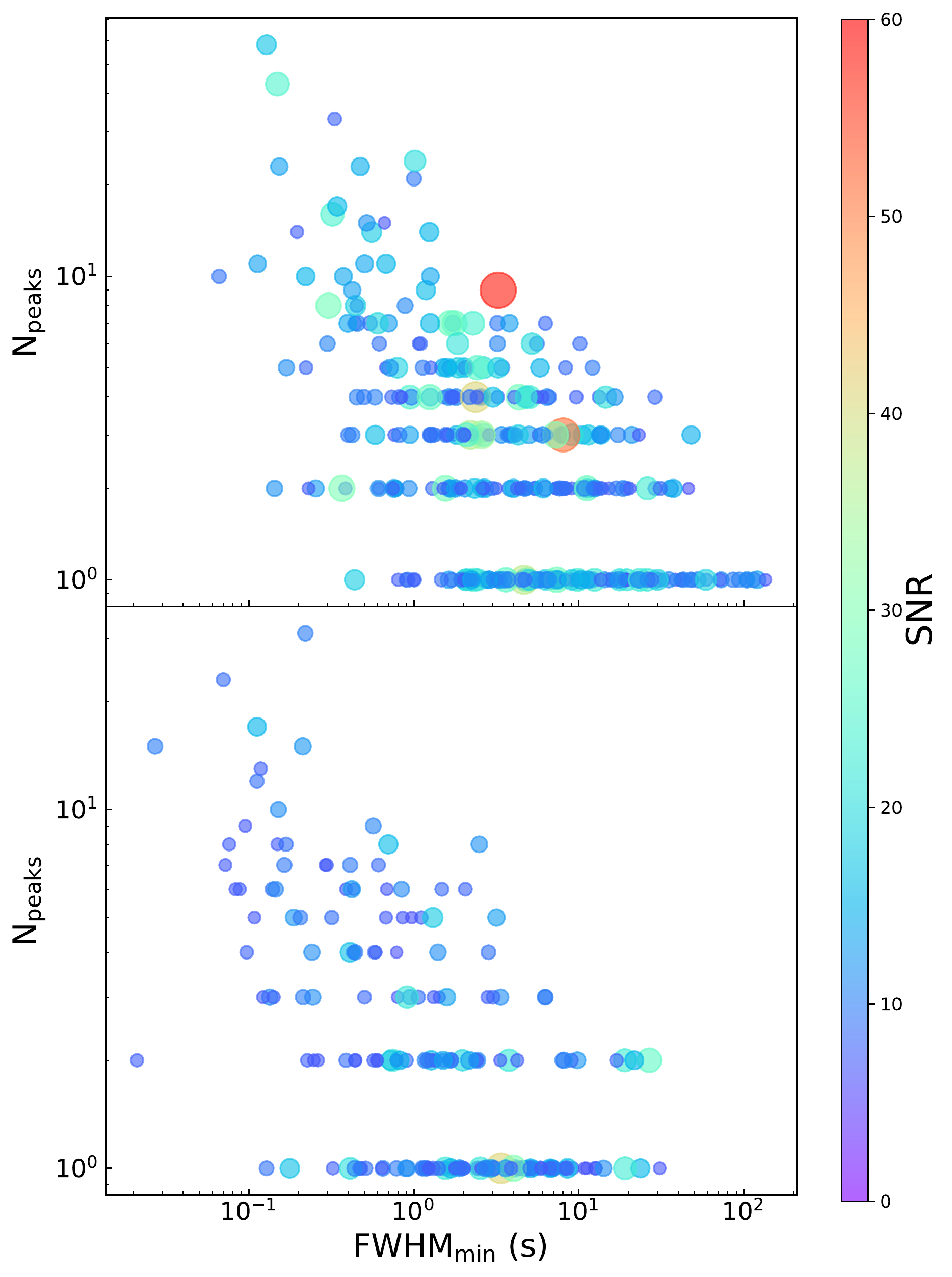}
\caption{Number of peaks vs. FWHM$_{\rm min}$, Type-II GRBs for the {\it Swift} sample (top panel; here only GRBs with known $z$ are shown for the sake of clarity) and {\it HXMT} sample (bottom panel). Symbol size and colour scale with SNR.}
\label{fig:Npeaks_FWHM_LGRBs}
\end{figure}

We investigated the role of the number of peaks within a GRB, as determined with {\sc mepsa}, in the FWHM$_{\rm min}$--$L_p$ correlation.
We therefore split the Type-II GRB sample with measured $z$ into four groups with comparable size, depending on the number of peaks that was previously established with {\sc mepsa}: $N_{\rm peaks}=1, 2, 3$~or~$4$, $\ge5$.
Left panel of Figure~\ref{fig:Lp_FWHM_differentNpeaks} shows the result.
GRBs with the largest number of pulses evidently cluster in the most luminous /narrow FWHM$_{\rm min}$ region. To understand whether this is mainly due to a SNR effect, we selected the 1/3 having the narrowest pulse with the highest SNR (threshold of SNR $>12.26$). The result is shown in the right panel of Fig.~\ref{fig:Lp_FWHM_differentNpeaks} and clearly proves that the conclusion remains unaffected.
%
To further characterise the link between number of peaks and FWHM$_{\rm min}$, in Figure~\ref{fig:Npeaks_FWHM_LGRBs} we show both observables in a scatter plot, where both colour and symbol size show the SNR. Notably, the FWHM$_{\rm min}$ significantly shifts towards lower values for GRBs with increasing number of pulses.

In principle, if the probability density function of the FWHM of a peak is the same for all GRBs, regardless of the number of peaks, a GRB with more peaks is more likely to have a small FWHM$_{\rm min}$, since more peaks means more trials. However, Figure~\ref{fig:Npeaks_FWHM_LGRBs} (top panel) clearly rules this case out: in fact, this putative distribution should result from sampling the $N_{\rm peaks}=1$ observed one, which is clearly incompatible with the FWHM$_{\rm min}$ distribution of GRBs with several peaks. We confirm this through the following test: we considered all the {\it Swift} Type-II GRBs (regardless of the knowledge of $z$) with $N_{\rm peaks}\ge10$ (hereafter, Obs10 sample). We then generated a population of fake GRBs (hereafter Sim10) with the same distribution of number of peaks as Obs10 (but with 100 times as many GRBs as Obs10), where each peak was chosen randomly from the distribution of GRBs with $N_{\rm peaks}=1$ (hereafter Obs1). We determined the FWHM$_{\rm min}$ distribution of Sim10 and compared with that of Obs10 through two-population KS and AD tests. The probability that the two distributions of FWHM$_{\rm min}$ are drawn from the same parent populations is $3\times 10^{-9}$ ($<10^{-3}$) with KS (AD) test. 
Using the same procedure with the {\it HXMT} sample, so including GRBs with unknown $z$ (bottom panel of Fig.~\ref{fig:Npeaks_FWHM_LGRBs}), we obtain a probability of $5\times 10^{-4}$ ($<10^{-3}$) according to the KS (AD) test.

\subsection{Lorentz factor vs. FWHM$_{\rm min}$}
\label{sec:FWHM_Gamma0}
To study the relation between MVT and Lorentz factor of the ejecta, we selected a subsample of Type-II GRBs for which an estimate of the latter was available in the literature. To this aim, we considered the following references: \citet{Lu12c,Yi17,Xue19b,Ghirlanda18}, along with a couple of additional references relative to individual GRBs: 111228A \citep{Xin16} and 150910A \citep{Xie20}.
All of these estimates, except for the GRBs of \citet{Xue19b} who obtained pseudo estimates based on the $E_{\rm p,i}-L_{\rm iso}-\Gamma_0$ relation \citep{Liang15}, are based on the peak time of the early afterglow, interpreted as the deceleration of the fireball in the thin shell regime. Because of this, for the GRBs in common we assumed the $\Gamma_0$ values obtained from the deceleration peak. In any case, we graphically distinguish the different sources, to spot any possible difference.
Our final sample consists of 131 GRBs, 39 of which have $\Gamma_0$ from the deceleration peak from \citet{Lu12c,Yi17,Xin16,Xie20} and 92 are pseudo values from \citet{Xue19b}.

We treated the sample of \citet{Ghirlanda18} separately, given that this is a rich collection that resulted from a homogeneous selection and treatment, and which also provides a set of lower limits on $\Gamma_0$ for many GRBs. Specifically, we calculated $\Gamma_0$ using their Equation~(11), taken from \citet{Nava13}. The estimated values were obtained for 50 GRBs that our sample shares with their golden and silver samples, while for the remaining 74 common GRBs we calculated the corresponding lower limits.
We checked the mutual consistency of the estimated values taken from \citet{Ghirlanda18} and the other references for a sample of common GRBs (41 values and 50 lower limits) and we found only a few incompatible estimates, whereas most estimates differ by $\lesssim20$\% and a dozen or so of lower limits that are incompatible with the estimates provided by the other works.

%
\begin{figure*}
\centering
\includegraphics[width=9 cm]{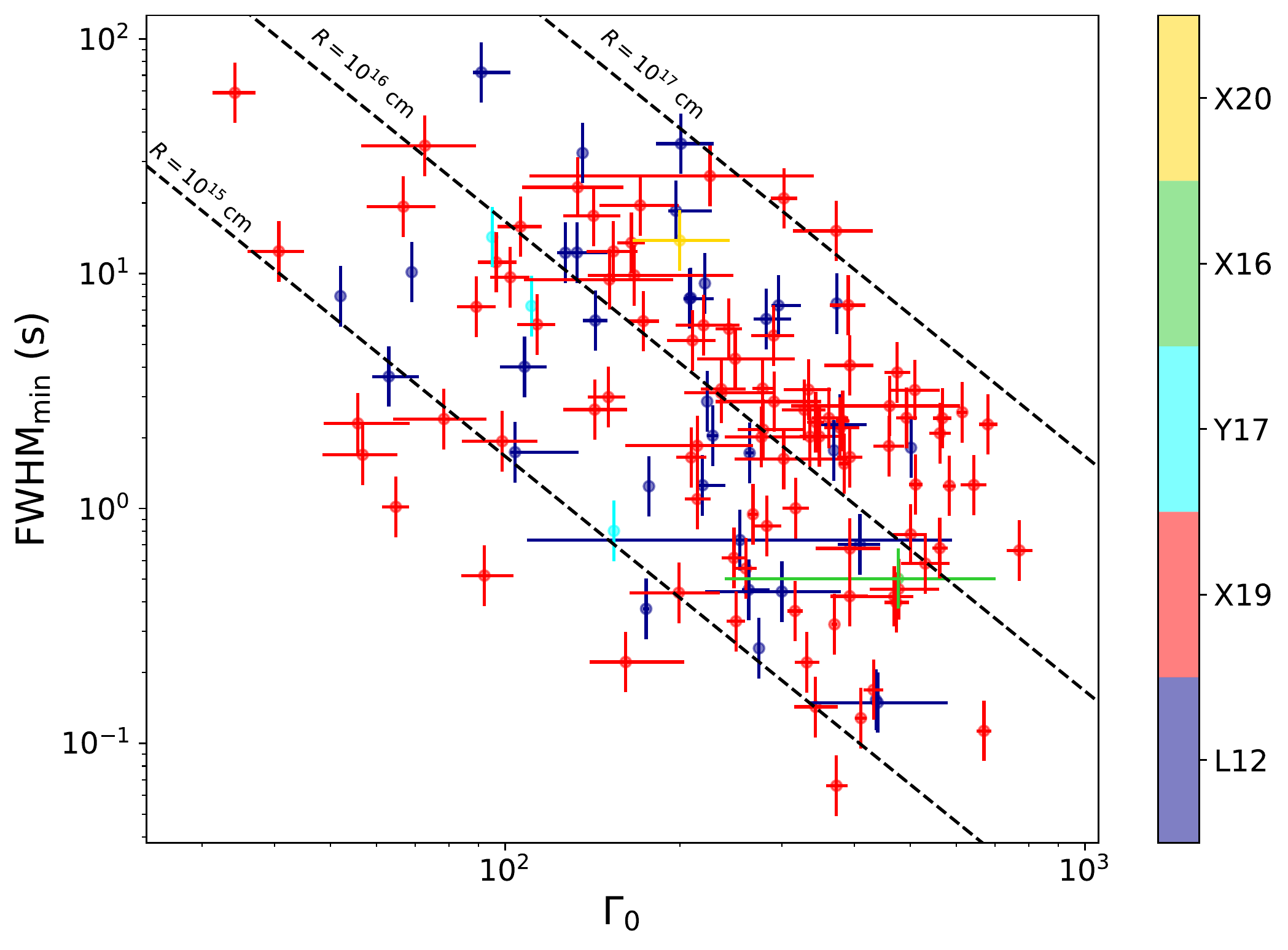}
\includegraphics[width=8 cm]{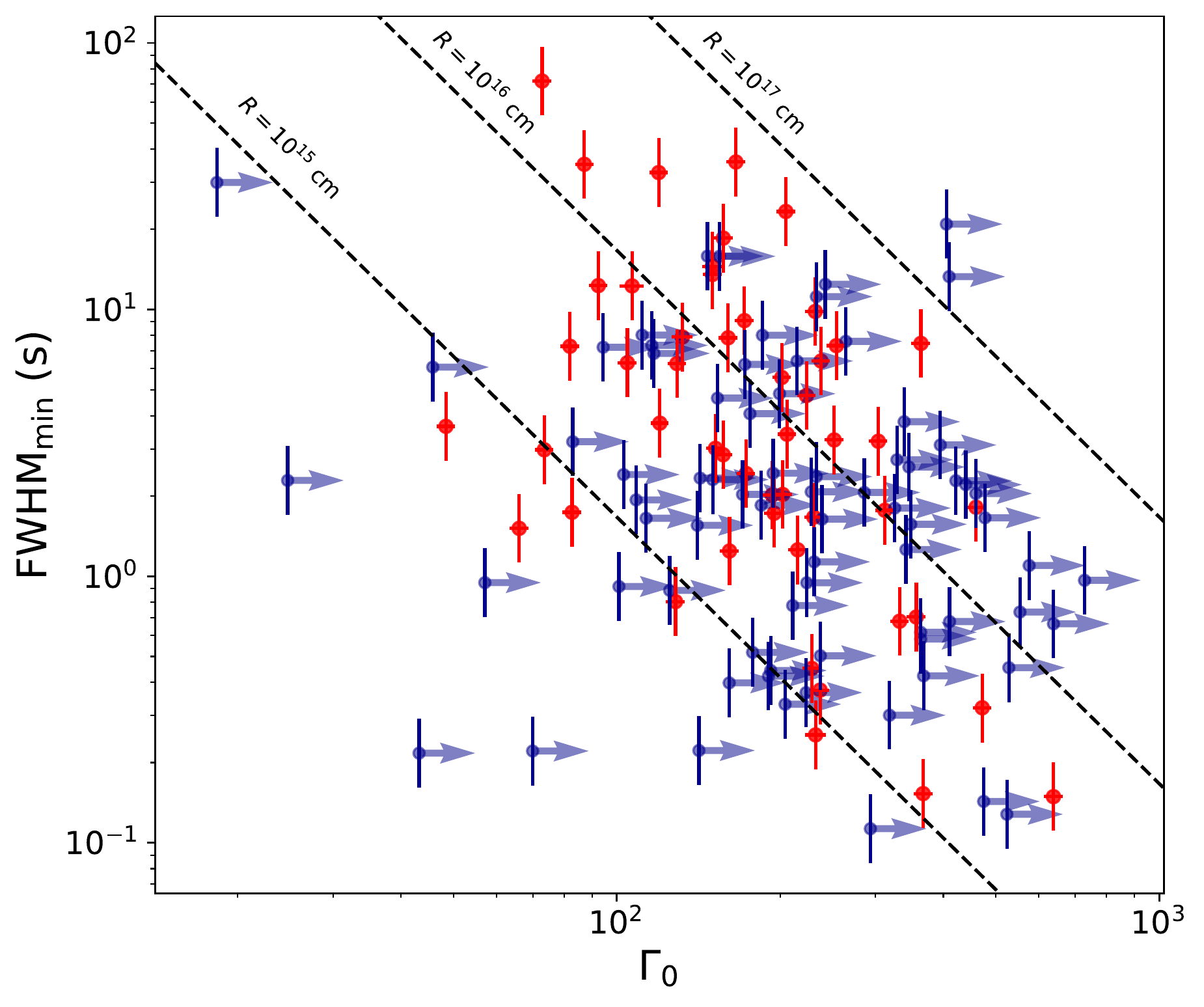}
\caption{{\it Left:} FWHM$_{\rm min}$ vs. the initial Lorentz factor $\Gamma_0$ for a number of Type-II GRBs. The values for the latter were taken from different data sets, which are colour-coded: L12 \citep{Lu12c}, X19 \citep{Xue19b}, Y17 \citep{Yi17}, X16 \citep{Xin16}, and X20 \citep{Xie20}. Dashed lines correspond to the typical distance $R= 2\,c\,\Gamma_0^2\,{\rm FWHM}_{\rm min}$ at which the dissipation process responsible for the prompt emission could take place. {\it Right}: same as left, except that we calculated $\Gamma_0$ from the data set by \citet{Ghirlanda18}. Red points are the GRBs belonging to their golden and silver samples, while blue points are lower limits on $\Gamma_0$.}
\label{fig:Gamma0_vs_FWHM}
\end{figure*}
%

Figure~\ref{fig:Gamma0_vs_FWHM} shows the results. The left panel shows the GRBs with colour-coded reference for the $\Gamma_0$ estimates, whereas the right panel uses the sample in common with \citet{Ghirlanda18}. The anti-correlation is evident for both data sets, although significantly scattered.

It is not straightforward to understand the impact of the selection bias affecting the sample of FWHM$_{\rm min}$ discussed in Section~\ref{sec:Lp_vs_FWHM}. Consequently, the apparent power-law slope of this correlation should be taken with care. Yet, interestingly both data sets show an overall consistency with FWHM$_{\rm min}\propto\Gamma_0^{-2}$, as shown by the dashed lines that correspond to constant values of $R = 2\,c\,\Gamma_0^2$~FWHM$_{\rm min}$, which is the typical distance at which the dissipation of energy into gamma-rays is supposed to take place. Almost all GRBs lie in the $10^{15} \lesssim R/{\rm cm} \lesssim 10^{17}$ range.

This range appears to be larger than $10^{13}$--$10^{14}$~cm, which is usually estimated in the case of internal shocks (e.g., \citealt{ZM04}). However, there are several cases for which the estimates are consistent with our results: for a number of {\it Swift} GRBs for which the early X-ray afterglow steep decay could be interpreted as high-latitude emission, the onset time of this decay enabled to obtain $R\ge 10^{15}$--$10^{16}$~cm \citep{Lyutikov06,Kumar07,LazzatiBegelman07,Hascoet12}. Also for the naked-eye burst 080319B it was found $R\gtrsim10^{16}$~cm, due to its bright prompt optical emission, which requires a synchrotron self-absorption frequency that is not too far above the optical band \citep{Racusin08,KumarPanaitescu08}.

Different models can accomodate these values for $R$: for a Poynting flux dominated outflow a typical value of $3\times10^{16}$~cm is predicted \citep{LyutikovBlandford03}. Alternatively, hybrid models like the Internal Collision-induced MAgnetic Reconnection and Turbulence (ICMART; \citealt{ICMART}) based on internal shocks, in which the dissipation mechanism is magnetic reconnection, also predict $R\sim10^{15}$--$10^{16}$~cm. Also classical multi-zone internal shocks with reasonable Lorentz factor distributions for the wind of shells can predict dissipation radii in the range $10^{15}$--$10^{16}$~cm, whereas the possible associated high-energy (HE) neutrinos and ultra-high energy cosmic rays (UHECRs) are mostly expected to be accelerated at distances $\sim10$--$100$ times closer to the engine \citep{Bustamante15,Bustamante17}. Our results may suggest that, if most GRB internal shocks take place above $\sim10^{15}$~cm, that would explain why bright GRB prompt emission is not accompanied by HE neutrinos, at least as long as low-luminosity GRBs are ignored. This is in agreement with the upper limit of 1\% on the contribution due to cosmological GRBs to the observed HE neutrino diffuse flux \citep{IceCube22_GRB}.

For each GRB we also compared the dissipation radius with the deceleration one $R_{\rm dec}$ at which the ejecta decelerates, which corresponds to the time at which the afterglow LC peaks. In particular, we took the $\Gamma_0$ and afterglow peak times of the sample by \citet{Ghirlanda18}, and verified that for all GRBs it is $R<R_{\rm dec}$.

\subsection{Jet opening angle vs. FWHM$_{\rm min}$}
\label{sec:FWHM_thetaj}
\begin{figure*}
\centering
\includegraphics[width=9 cm]{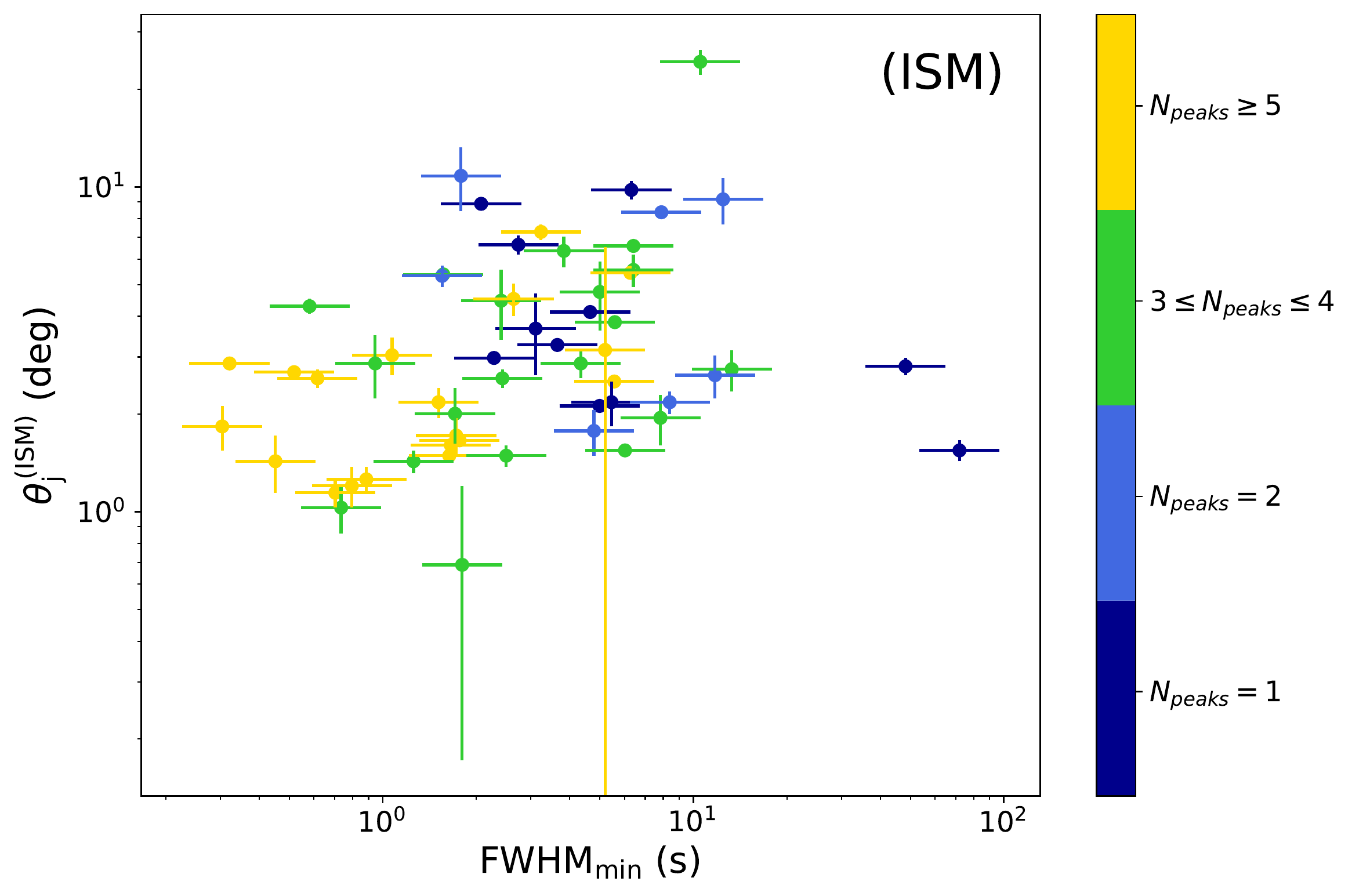}
\includegraphics[width=9 cm]{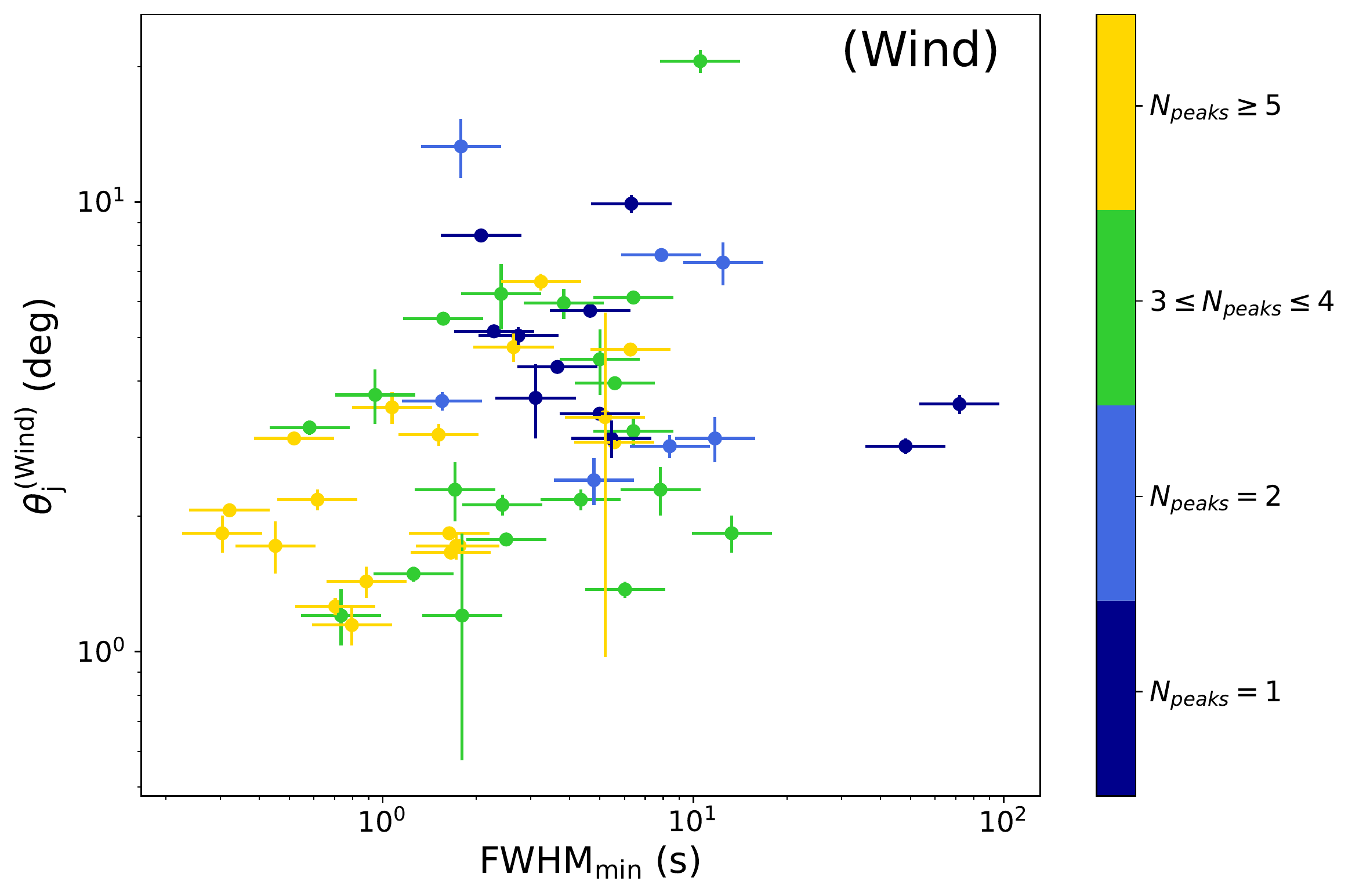}
\caption{Jet opening angles vs. FWHM$_{\rm min}$ for a set of Type-II GRBs for which both quantities could be estimated, assuming two different circumburst environments: ISM (left), wind (right). Colors indicate the number of pulses.}
\label{fig:thetaj_vs_FWHM}
\end{figure*}

We explore the relation between MVT and jet opening angle $\theta_j$ as measured from the afterglow modelling for the GRBs for which evidence for an achromatic jet break was found. To this aim, we took the values reported in \citet{Zhao20} in the two cases of ISM and wind environments. The resulting subset with measured $\theta_j$ includes 57 of our Type-II GRB sample with measured FWHM$_{\rm min}$ and number of peaks. Figure~\ref{fig:thetaj_vs_FWHM} shows the results for ISM (left panel) and wind (right panel): the information on the number of pulses is colour-coded. 
%

Despite the apparent scattering, there is evidence for a correlation between jet opening angle and MVT with significance values of a few $10^{-3}$ (Table~\ref{tab:thetaj_FWHM_corr}).
\begin{table}[]
    \centering
    \begin{tabular}{l|l|c}
    \hline
    Environment   &  Test  & p-value\\
     \hline
    ISM  & Pearson  & $2.4\times10^{-2}$\\
    ISM  & Spearman & $5.6\times10^{-3}$\\
    ISM  & Kendall  & $5.2\times10^{-3}$\\
    Wind  & Pearson  & $4.6\times10^{-3}$\\
    Wind & Spearman & $2.0\times10^{-3}$\\
    Wind & Kendall  & $4.4\times10^{-3}$\\
\hline
    \end{tabular}
    \caption{P-values of tests of correlation between MVT and jet opening angle for a sample of 57 Type-II GRBs displayed in Figure~\ref{fig:thetaj_vs_FWHM}.}
    \label{tab:thetaj_FWHM_corr}
\end{table}
Motivated by the result in Fig.~\ref{fig:thetaj_vs_FWHM} we further tested the apparent clustering of GRBs with numerous peaks in the narrow jet--short MVT through a 2-D two-population KS test \citep{NumRecC}, after dividing the sample into two groups, one including 18 GRBs with $N_{\rm peaks}\le 2$ and another with 39 GRBs having $N_{\rm peaks}>2$. Using the ISM (wind) values of $\theta_j$ the probability that the number of peaks is uncorrelated with the position in the FWHM$_{\rm min}$--$\theta_j$ plane is $4.1\times10^{-3}$ ($1.5\times10^{-3}$), equivalent to $2.9 \sigma$ ($3.2 \sigma$).

\begin{table*}[]
    \centering
    \begin{tabular}{l|c|c|c|c|c|c}
    \hline
    \hline
GRB Name & $L_{\rm p}$ ($10^{50}$ erg/s)& $\Gamma_{0}$ & Ref. &     $ \Gamma_{0}^{(G)}$ &     $\theta_{j}^{\rm ISM}$ (rad) &    $\theta_{j}^{\rm Wind}$ (rad)\\

\hline
\hline
 050126 & $4.13 \pm 0.66$ & - & - & - & - & -  \\
 050219A &$0.47 \pm 0.04$ & - & - & - & - & - \\
 050223 &$0.58 \pm 0.08$ &- & - & - & - & - \\
050315 &$37.4 \pm 3.0$ &- & - & -   &$0.111 \pm   0.012$&    $0.104 \pm 0.008$\\
 050318 &$38 \pm 4$ & - &- & $57 \pm 2$ & - & -\\
050319 &$95 \pm 15$ &- & - & -   &  $0.053 \pm   0.007$ & $0.061 \pm 0.005$\\
 050401 & $912 \pm 76$ &     $254_{-145}^{+335}$ & L12 & $554 \pm 25$ &   $0.018 \pm 0.003$ &   $0.021 \pm 0.003$\\
 050416A & $5.44 \pm 0.85$ &     -     &-     &       $101 \pm 6$ &  - &   -\\
 050505 &$311 \pm 40$ & - & - & - & - & - \\
 050525A &$69.4 \pm 1.5$ &     $268 \pm 5$ &    X19  &    $224 \pm 10$ &   $0.050 \pm 0.011$ &    $0.065 \pm 0.009$\\
 \hline
   
\label{tab:BAT2sample}
\end{tabular}
\caption{The first 10 GRBs of {\it Swift} sample; $ \Gamma_{0}^{(G)}$ is the Lorentz factor from \citet{Ghirlanda18} using Equation~(11) therein from \citet{Nava13}. This table is available in its entirety in machine-readable form.}
\end{table*}


%

\section{Discussion}
\label{sec:disc}
Our results show that the observation of ms-long individual pulses is very rare in either class of GRBs. On the one end, even if they exist, their identification is limited by instrumental sensitivity, given the strong dependence of the detection threshold on FWHM$_{\rm min}$ (Fig.~\ref{fig:PR_FWHM}).  On the other end, we do observe rare GRBs with a peak rate high enough to ensure the possibility to detect ms-long pulses, but in practice this seems to be rarely the case. The short end of the MVT distributions of both Type-I and Type-II GRBs extends to $\sim10$~ms within a factor of a few, depending on the energy passband.
The two classes have overlapping MVT distributions, although the populations are statistically different. In this respect, SEE-GRBs, i.e. Type-I GRBs with an ambiguous time profile, exhibit an average MVT that is shorter than that of Type-II GRBs, making our estimate of MVT a useful indicator for the progenitor class. In SEE-GRBs, whose environmental properties are indistinguishable from those of SGRBs \citep{Fong22,Nugent22}, the presence of a prolonged $\gamma$--ray activity characterised by a longer MVT that follows an initial spike could hint to a long-lived spinning down protomagnetar \citep{Metzger08,Bucciantini12}, a scenario which can also account for the complex broadband evolution exhibited by GRBs like 180618A \citep{Jordana22}. There are alternative interpretations, such as the one proposed to explain 211211A, in which the SEE would be produced by the prolonged accretion-powered activity of a newly formed black hole (BH), ruled by the strong magnetic field of one of the merging NSs \citep{Gao22}.
Regardless of the possible presence of the SEE and of its interpretation, the shorter MVT of Type-I GRBs appears to be a distinctive property with respect to the bulk of Type-II events. 
Although the nature of the inner engines that power the two classes could be similar (e.g., either a supramassive ms-magnetar or an hyperaccreting newly formed BH), the longer MVT in Type-II GRBs might be due to the engine variability being smeared out by the interaction with the massive envelope, whereas in Type-I the central engine variability imprint in the jet is retained throughout the propagation in less massive merger ejecta (e.g., see \citealt{Gottlieb22a}).

The other major difference between the two classes is the apparent absence of correlation between MVT and peak luminosity for Type-I GRBs, as shown in Fig.~\ref{fig:Lp_FWHM_BAT_LGRBs_sim}.

Concerning the properties of Type-II GRBs with measured redshifts, upon a careful evaluation of the involved selection effects on both luminosity and FWHM$_{\rm min}$, we have confirmed that $L_{\rm p}$ and MVT do correlate, although a mathematical description that is corrected for the selection effects requires more extensive simulations that go beyond the scope of the present investigation. Additional information comes from the other correlations that we report here and involve MVT: in particular, (Type-II) GRBs that display many ($>2$) pulses, on average have short MVT (typically in the range $0.01$--$1$~s), are more luminous, have a higher bulk Lorentz factor, and, whenever the information is available, tend to have narrower jets or a smaller off-axis angle.

The jet opening angle $\theta_j$ is usually estimated from afterglow modelling and, in particular, from the observation of an achromatic break in the afterglow light curves that would correspond to the time at which the Lorentz factor $\Gamma$ of the forward shock is such that $1/\Gamma\sim\theta_j$. While this is true for an on-axis observer, for $\theta_{\rm obs}\not =0$ but $\theta_{\rm obs}<\theta_j$, the jet break time is actually set by the furthest edge from the observer, i.e. when the relativistic beaming decreases to the point at which $1/\Gamma\sim \theta_j+\theta_{\rm obs}$ \citep{Vaneerten10a}. In practice, when one also includes other factors, such as jet spreading, angular structure of a jet (as opposed to the simplistic case of a top-hat jet), radial fluid structure, and arrival time effects, deviations of the afterglow LC from simple power laws hinder an accurate measure of both jet and observer angles \citep{Vaneerten12c}. Besides, for increasing observer angles, but still $\theta_{\rm obs}<\theta_j$, the jet break time may occur correspondingly later by a factor of 3--5 \citep{DeColle12b}. Consequently, the variable that we found to correlate with MVT is likely more indicative of $(\theta_j+\theta_{\rm obs})$ or, at least, of a combination of both $\theta_j$ and $\theta_{\rm obs}$ rather than $\theta_j$ alone for an on-axis view ($\theta_{\rm obs}\simeq0$).

A simple interpretation of these correlations invoke a structured jet viewed through a range of different observer angles for different GRBs. Simulations suggest that the angular structure of a GRB jet consists of a flat core with an opening angle $\theta_j$, followed by a power-law decline ($E_{\rm iso}\propto\theta^{-\delta}$ for $\theta_j<\theta<\theta_c$) which models the so-called jet-cocoon interface (JCI). This is an interface layer between the jet core and the cocoon and which extends to $\theta_c$ \citep{Gottlieb21a}. The power-law index $\delta$ depends on the jet magnetisation: $\delta\sim3$ for a weakly magnetised jet, $\delta\sim1$--$2$ for a purely hydrodynamic jet. This difference arises from the different degree of mixing between jet and cocoon at the JCI, which in turn affects the baryon loading of jet: a magnetised jet, whose existence is also supported by early-time optical polarisation measurements (e.g. \citealt{Gomboc08,Mundell13,Japelj14,Steele17}), would suffer from a reduced mixing, with less energy transferred to the JCI and a consequent steeper energy angular profile \citep{Gottlieb20b}.

Assuming similar values for $\theta_j$ for different GRBs as could be plausible in a quasi-universal jet structure (e.g., see \citealt{Salafia20} and references therein), or at least assuming that the spread of values of $\theta_{\rm obs}$ is greater than that of $\theta_j$ for the observed population, the variety of values measured from the afterglow LC for the jet opening angle, which is actually more revealing of $(\theta_j+\theta_{\rm obs})$, mostly reflects the range of $\theta_{\rm obs}$. For a relatively on-axis view, observed high Lorentz factor, high isotropic-equivalent peak luminosity are naturally accounted for. A short MVT and the abundance of pulses would suggest that we are looking through the jet core at the inner engine activity, unaffected by smoothing and blending that instead would characterise a more off-axis view, but still close to the JCI boundary ($\theta_{\rm obs}\sim\theta_j)$, because of lower Doppler boosting and longer arrival time delays (e.g., see \citealt{Salafia16}). In addition, the LC blending could hinder the identification of distinct adjacent pulses, thus explaining why these GRBs show fewer pulses on average. Concerning more off-axis GRBs, i.e. those with $\theta_j<\theta_{\rm obs}<\theta_c$, they would appear as low-luminosity GRBs, which are mostly missing from present GRB catalogues.

A possible additional key property is suggested by recent 3-D GRMHD simulations of the temporal evolution of a jet that results from a collapsar: some shocked gas would free fall towards the newly formed BH and would be then deflected by the jet toward the accretion disc, which would consequently make the jet tilt randomly with respect to the BH rotational axis \citep{Gottlieb22c}. The resulting jet would break out of the photosphere with a typical opening angle of $\theta_j\sim6^\circ$ and wobbling around with an angle $\theta_t\sim12^\circ$ and with hybrid composition of magnetic and thermal energy due to the variable mixing. In this picture, short MVT ($0.01$--$0.1$~s) would reflect the inner engine activity, possibly due to random fluctuations in the accretion process and launching mechanism, whereas long (1--10~s) interpulse times would occur whenever the jet points away from the observer \citep{Gottlieb22b}. While the possibility of precessing jets in GRB sources is not new \citep{PortegiesZwart99,PortegiesZwart01,Fargion01,Reynoso06,Lei07}, the stochastic nature of this wobbling, supported by state-of-the-art simulations, suggests that a multi-pulsed, luminous GRB with short MVT and high Lorentz factor could be due to a jet wobbling around the line of sight, thus giving rising to more short pulses and on average exposing the variability of the inner engine during the frequent on-axis alignments. Less luminous GRBs, with fewer pulses and longer MVT would harbour wobbling jets on average more misaligned with respect to the line of sight: these GRBs would correspond to the cases in which the angle between the time-average pointing direction of the jet and the line of sight is similar to the jet opening angle, $\langle\theta_{\rm obs}\rangle\sim\theta_j$. This would turn into fewer pulses, given that the jet would spend more time with the line of sight off the jet core. The scatter of the observed correlations could be explained as due to the fact that even for such a GRB the probability of a temporary fortuitous on-axis view, thus giving a narrow pulse, is not negligible, although less likely.
This interpretation is supported by some Monte Carlo simulations of randomly wobbling structured jets: using a different definition of variability related to the number of pulses, a connection is predicted between variability and the jet opening angle as measured from the afterglow light curves \citep{Budai20}, which agrees with our result.

A similar correlation between $\theta_j$ (or $\theta_j+\theta_{\rm obs}$, as explained above) and a former definition of variability was already established by \citet{Kobayashi02} for a few GRBs with measured quantities available at the time. According to their interpretation, GRBs with constant energy may result in different jet masses, so that a smaller mass loading would be associated with a narrower jet, higher Lorentz factor and isotropic-equivalent luminosity, as supported by simulations.

\begin{figure*}
\centering
\includegraphics[width=10cm]{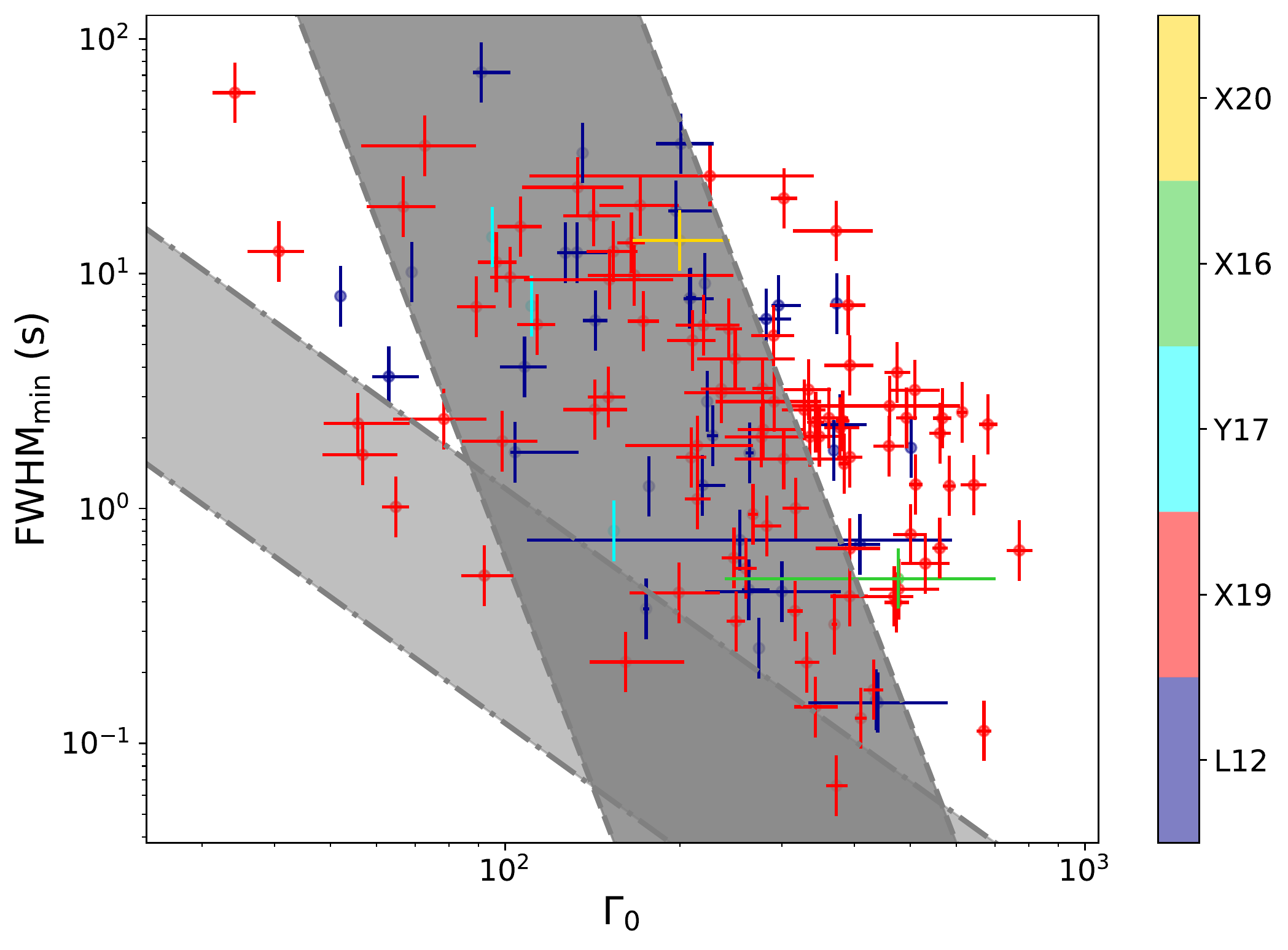}
\includegraphics[width=7.5cm]{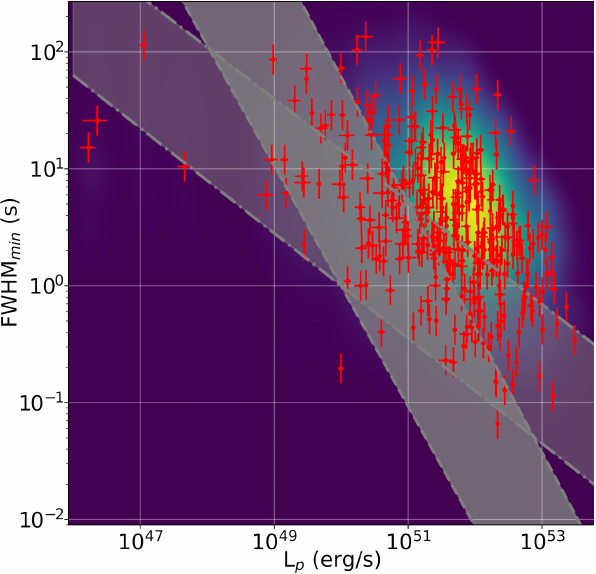}
\caption{{\it Left}: same as Figure~\ref{fig:Gamma0_vs_FWHM} along with the expectations discussed by \citet{Xie17}, highlighted by the light-shaded region between dash-dotted lines ($\nu\bar{\nu}$ annihilation mechanism) and by the dark-shaded region between dashed lines (BZ mechanism). {\it Right}: same as Fig.~\ref{fig:Lp_FWHM_BAT_LGRBs_sim}, except that the two axes are swapped for ease of comparison with the expectations discussed by \citet{Xie17}; we also show the expected regions from the same models as in the left panel.}
\label{fig:2mechanisms}
\end{figure*}

The scaling between MVT vs. Lorentz factor $\Gamma_0$ and MVT vs. luminosity may help constrain the mechanism used by the inner engine to power the relativistic jet. In this respect, \citet{Xie17} considered two alternative scenarios that invoke a BH: (i) extraction of the BH rotational energy through the magnetic field sustained by the accretion disc according to the Blandford \& Znajek mechanism (BZ;  \citealt{BlandfordZnajek77}); (ii) a neutrino-dominated accretion flow (NDAF; \citealt{Popham99}), whose neutrinos and anti-neutrinos annihilate and power the jet. In the BZ mechanism there is also a NDAF that supports and influences the magnetic field, which in turn suppresses the baryon loading from the neutrino-driven wind. In either case the MVT is determined by the viscous instability of the NDAF.
In Figure~\ref{fig:2mechanisms} we show our results on the correlations between MVT and $\Gamma_0$ (left panel) and between MVT and peak luminosity (right panel), along with the regions expected in each of the two scenarios: the light-shaded region between dash-dotted lines refers to the $\nu\bar{\nu}$ annihilation mechanism, whereas the dark-shaded region between dashed lines refers to the BZ one. The values of each set of model parameters are the same as the ones that are considered in \citet{Xie17}: as a consequence, while both regions can be shifted parallely, their slopes are more of a distinctive property of each and, as such, should be considered. While the slopes of the MVT--$\Gamma_0$ predicted relations seem overall compatible with the data set, which is significantly scattered, the same comparison in the MVT--$L_{\rm p}$ plane seems to favour the BZ mechanism, as also argued by \citeauthor{Xie17}: the $\nu\bar{\nu}$ annihilation model looks significantly  shallower (FWHM$_{\rm min}\propto (L_{\rm p}^{(\nu\bar{\nu})})^{-1/2}$) than the data, whereas the BZ slope is FWHM$_{\rm min}\,\propto\,(L_{\rm p}^{\rm (BZ)})^{-1}$.
%
%

The relations found for a common sample of GRBs and blazars by \citet{Wu16}, MVT$\propto\Gamma_0^{-4.7\pm 0.3}$ and MVT$\propto L_{\rm p}^{-1.0\pm 0.1}$ are both compatible with our samples, although the large scatter observed in our MVT--$\Gamma_0$ sample makes it noncritical.

An hyperaccreting Kerr BH with a magnetised torus is expected to power via BZ a magnetically dominated jet. MHD simulations identify the BH spin as the main driver of the GRB variability along with the magnetic field strength, although other factors should be considered that contribute to the conversion efficiency of the dissipation of energy into gamma--rays (e.g., \citealt{Granot15rev}). The MVT, in terms of minimum duration of pulses, would reflect the timescale of the fastest growing mode of magneto-rotational instabilities in the accretion disc \citep{Janiuk21}. In this scenario, a higher BH spin would launch a jet with a correspondingly higher Lorentz factor and shorter MVT, thus accounting for the observed correlation. On average, for small values of $\theta_{\rm obs}$ a higher $\Gamma_0$ is also expected, although the relation does not appear always to be monotonic for different values of BH spin parameter and magnetic field strength.

%

\section{Conclusions}
\label{sec:conc}
We proposed a simple definition of MVT of GRB prompt emission as the FWHM of the shortest pulse that is identified through {\sc mepsa}, a thoroughly tested GRB peak search algorithm. We applied this method to two independent and complementary GRB data sets: {\it Swift}/BAT and {\it Insight-HXMT}/HE, both of which were split into two groups: Type-I and Type-II GRBs, the former including SEE-GRBs. Firstly, ms-long MVT is very rarely observed, the shortest values being around 10~ms. However, this could be partly due to a SNR-related selection effect. Although the two groups have overlapping MVT distributions, MVT of Type-I GRBs is in the range 10~ms--1~s and is on average significantly shorter than that of Type-II GRBs, which encompasses the range $\sim0.1$--$100$~s. Remarkably, also SEE-GRBs with $T_{90}>2$~s and characterised by deceptively long and structured time profiles such as 060614 and 211211A, display a short MVT that is more typical of Type-I GRBs, thus propelling this definition of MVT as a useful indicator of the progenitor class, especially in the presence of ambiguous $\gamma$--ray time profiles. The origin of this difference could stem from the different ejecta masses that the relativistic jets of either class have to pierce through.

Concerning the subsample of Type-II GRBs with measured redshift, upon a careful evaluation of the selection effects that impact the MVT measure in the MVT-peak rate plane, we confirm the existence of anti-correlations between MVT and peak luminosity $L_{\rm p}$, and between MVT and initial Lorentz factor of the ejecta $\Gamma_0$. Moreover, we could establish that MVT also correlates with the number of peaks and the jet opening angle (measured from the achromatic break in the afterglow light curves), although the latter is probably more of a sum of jet and observer angles. Taken together, we find that GRBs with short MVT ($0.1$--$1$~s) on average have narrower jets and/or smaller observer angles ($\lesssim2$--$4^\circ$), higher Lorentz factors ($\Gamma_0\gtrsim100$), high peak luminosities ($L_{\rm p}\gtrsim10^{51}$~erg~s$^{-1}$) and exhibit several pulses.
A possible interpretation that builds on 3-D GRMHD state-of-the-art simulations of a jet propagating through stellar envelopes, involves a structured jet with a flat core ($0\le\theta<\theta_j$) and a power-law profile that models the jet-cocoon interface ($\theta_j<\theta<\theta_c$). GRBs with short MVT would be seen within the jet core, resulting in higher $L_{\rm p}$ and $\Gamma_0$, shorter and more numerous peaks, and revealing the true variability imprinted by the inner engine, such as an hyperaccreting BH possibly powered via the Blandford-Znajek mechanism. Conversely, GRBs viewed across the boundary between the jet core and the jet-cocoon interface, would appear as less luminous, with lower Lorentz factors and longer MVT due to a smaller Doppler boosting and longer arrival time delays. The possibility that such a jet could wobble randomly within angles comparable if not greater than the jet core itself, further suggests that the different number of peaks observed in  different GRBs could indicate how often the jet core points to the observer, thus revealing how off axis the line of sight is with respect to the average jet direction.

\begin{acknowledgements}
We are grateful to the Referee for their useful comments which helped us to improve the paper. AEC and CG acknowledge financial support from FIRD 2022 of UNIFE Dept. Physics and Earth Science under the project ``Caratterizzazione e simulazioni di curve di luce di Gamma-Ray Burst come processi stocastici e validazione attraverso tecniche di machine learning'' (PI: CG). CGM and NJM acknowledge support from Hiroko and Jim Sherwin. This work is supported by the National Program on Key Re-search and Development Project (2021YFA0718500) and the National Natural Science Foundation of China under grants 11733009, U1838201 and U1838202. This work made use of data from the {\it Insight}-HXMT mission, a project funded by China National Space Administration (CNSA) and the Chinese Academy of Sciences (CAS).
\end{acknowledgements}






\begin{appendix}
\section{Estimating the peak FWHM through {\sc mepsa}}

{\sc mepsa} (Multiple Excess Peak Search Algorithm) is an algorithm aimed at identifying peaks in LCs affected by uncorrelated Gaussian noise. {\sc mepsa} scans the time series at different timescales by comparing a peak candidate with a variable number of adjacent bins; the number of adjacent bins involved in the detection is called $N_{\rm adiac}$. 

When the entire LC has been screened, the process is re-run on the rebinned versions of the same curve; each time the rebinning factor is increased by one up to a maximum established by the user.
At the end of the procedure, {\sc mepsa} provides the detection timescale $\Delta t_{\rm det}$ for each peak candidate: it is the product between the original time resolution of the time series and the rebinning factor. We refer the reader to \citet{Guidorzi15a} for a more detailed description. 

Since {\sc mepsa} does not provide direct information on the FWHM of a detected peak, we had to preliminarily calibrate it, starting from the rule of thumb declared in \citet{Guidorzi15a}: this establishes a set of different scalings between $\Delta t_{\rm det}$ and FWHM for different ranges of SNR. In order to determine a more precise calibration between FWHM, $\Delta t_{\rm det}$, and SNR, we simulated LCs with FWHM values taken from a given lognormal distribution. For our calibration we used 1600 Fast Rise and Exponential Decay (FRED) profiles:

\begin{equation}
  F(t)=\begin{cases}
    A \exp{\Big[-\Big(\frac{t_0-t}{t_{r}}\Big)^{p}\Big]} , & \text{if $t<t_{0}$}\\
   A \exp{\Big[-\Big(\frac{t-t_{0}}{t_{d}}\Big)^{p}\Big]} , & \text{if $t\ge t_{0}$}\;
  \end{cases}
\end{equation}

with $t_{0}=0$, $p=1.5$, $t_{d}=3t_{r}$, FWHM$=10^{x}$~s, where $x$ is random normal distributed with $\mu=\log{0.6}$ and $\sigma=0.5$, following the same prescriptions adopted in \citet{Guidorzi15a}. 
Simulated LCs at 1 ms were rebinned with rebinning factor = 1, 4, 64, 1000. These profiles have finally been affected by uncorrelated Gaussian noise.

\begin{figure}
\centering
\includegraphics[width=9 cm]{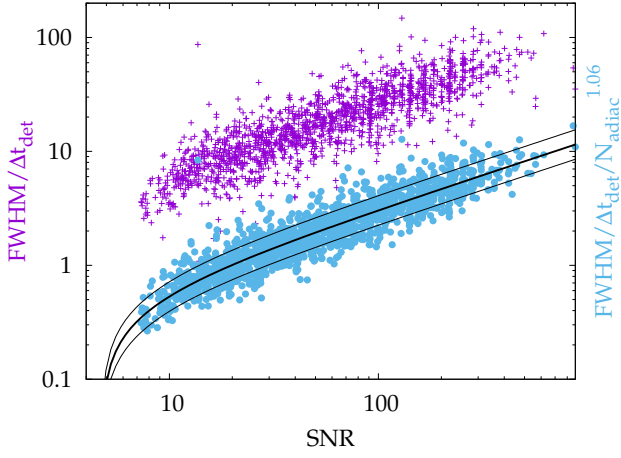}
\caption{Ratio between true FWHM and {\sc mepsa} detection timescale $\Delta\,t_{\rm det}$ (red crosses; left-hand side vertical axis) vs. SNR for a sample of simulated GRB-like pulses. The right-hand side vertical axis displays the same ratio further divided by {\sc mepsa} parameter $N_{\rm adiac}^{1.06}$ for the same data set (blue circles). The latter quantity is found to minimise the scatter around the best fitting relation.}
\label{fig:MEPSA_calib}
\end{figure}

We tried to include further parameters provided by {\sc mepsa} to see whether we could further reduce the scatter around the best fittig relation. To this aim, we assumed that the FWHM could be described by the following relation:
\begin{equation}
\rm FWHM\ \propto\ \Delta t_{\rm det}\ \left(\frac{\rm SNR}{{\rm SNR}_{0}}-1\right)^\alpha \ \rm P^\beta\;,
\label{eq:FWHMeq1}
\end{equation}
where SNR$_{0}$ is a hard lower limit for SNR that had preliminarily been fixed to $4.7$, $P$ is a generic {\sc mepsa} parameter and $\alpha$ and $\beta$ are power-law indices to be determined.
After choosing $P$, we considered the corresponding logarithmic quantities of the multiplicative terms in eq.~(\ref{eq:FWHMeq1}) and determined $\alpha$ and $\beta$ by finding the maximum likelihood within a linear model regression approach (see Section~3.1.1 of \citealt{BishopML}). As a result, we found a significant improvement by using $N_{\rm adiac}$ as the third parameter $P$. As shown in Figure \ref{fig:MEPSA_calib}, the best fitting relation correlates FWHM with $\Delta t_{\rm det}$, SNR and $N_{\rm adiac}$:
\begin{equation}
\begin{split}
\rm FWHM &=\ 10^{-0.31}\ \Delta t_{\rm det}\  \ \left(\frac{\rm SNR}{4.7}-1\right)^{0.60} \ \rm N_{\rm adiac}^{1.06}\\
\sigma_{\rm FWHM} &=\ \rm FWHM \ (10^{\pm0.13} - 1)
\end{split}
\label{eq:FWHMeq2}
\end{equation}
The scatter around the best fitting relation, expressed by $\sigma_{\rm FWHM}$, is calculated as the multiplicative coefficient that corresponds to the logarithmic standard deviation of the simulated points. To account for the asymmetric nature of the corresponding uncertainty on FWHM, in eq.~(\ref{eq:FWHMeq2}) we distinguish between negative and positive uncertainties, but in practice the $1$--$\sigma$ uncertainty on FWHM that affects the calibration of eq.~\ref{eq:FWHMeq2} is about 35\%. Consequently, all the FWHM values and relative uncertainties that were derived through {\sc mepsa} in this paper have been calculated using eq.~(\ref{eq:FWHMeq2}).

%

\end{appendix}


\end{document}